\documentclass[12pt]{article}
\pdfoutput=1
\usepackage[letterpaper, margin=1in]{geometry}
\usepackage{amsmath, amssymb, mathrsfs, amsfonts, cite, graphicx, slashed, cancel, bm, verbatim, tabularx, pict2e, multirow, subfigure, hhline, braket, float, stackengine, setspace, authblk, physics, bbm, dsfont, etoolbox, booktabs}
\usepackage[font=small,labelfont=bf]{caption}
\usepackage[normalem]{ulem}

\allowdisplaybreaks

\usepackage[utf8]{inputenc}

\usepackage[dvipsnames]{xcolor}
\RequirePackage{luatex85}
\usepackage{array} 
\usepackage[super]{nth}

\pagecolor{white}

\stackMath

\definecolor{DarkOrange}{rgb}{0.99,0.5,0}
\definecolor{DarkGreen}{rgb}{0,0.6,0}
\definecolor{goodgreen}{rgb}{0,.6,0.4}

\let\OLDthebibliography\thebibliography
\renewcommand\thebibliography[1]{
	\OLDthebibliography{#1}
	\setlength{\parskip}{3.0pt plus 2.5pt minus 1.0pt}
	\setlength{\itemsep}{3.0pt plus 2.5pt minus 1.0pt}
}

\usepackage[colorlinks=true,linktocpage=true,linkcolor=DarkOrange,citecolor=DarkGreen,urlcolor=DarkGreen]{hyperref}

\newcommand{\losigma}{[\mathcal{S}_1]}
\newcommand{\nlosigma}{[\mathcal{S}_2]}
\newcommand{\lodelta}{[\mathcal{D}_1]}
\newcommand{\nlodelta}{[\mathcal{D}_2]}
\newcommand{\eV}{\,\text{eV}}
\newcommand{\keV}{\,\text{keV}}
\newcommand{\gev}{\,\text{GeV}}
\newcommand{\mpl}{M_\text{Pl}}

\newcommand{\Trh}{T_{\rm RH}}
\newcommand{\xrh}{x_{\rm RH}}
\newcommand{\M}{\mathcal{M}}
\newcommand{\Hub}{\mathbf{H}}
\newcommand{\Hrh}{\mathbf{H}_{\rm RH}}
\newcommand{\beq}{\begin{eqnarray}}
\newcommand{\eeq}{\end{eqnarray}}
\newcommand{\nnmb}{\nonumber}
\newcommand{\lrf}[2]{\left(\frac{#1}{#2}\right)}
\newcommand{\lag}{\mathscr{L}}
\newcommand{\ccdot}{\!\cdot\!}
\newcommand{\Neff}{N_\text{eff}}

\newcommand{\CMB}{\text{CMB}}

\usepackage{cleveref}
\crefname{table}{Tab.}{Tabs.}
\crefname{equation}{Eq.}{Eqs.}
\crefname{eqnarray}{Eq.}{Eqs.}
\crefname{appendix}{Appx.}{Apps.}
\crefname{section}{Sec.}{Secs.}
\crefname{figure}{Fig.}{Figs.}

\title{Genesis of Baryon and Dark Matter Asymmetries \\ through Ultraviolet Scattering Freeze-in}

\author{Pouya Asadi$^1$, Marianne Moore$^{2,3}$, David E. Morrissey$^4$, Michael Shamma$^{4,5}$ \\
{\small \color{purple} 
\texttt{pasadi@uoregon.edu, mamoore@mit.edu, dmorri@triumf.ca, mshamma@triumf.ca},}
\\\vskip1em
{\small \textit{${}^1$Institute for Fundamental Science and Department of Physics,}\\
\textit{University of Oregon, Eugene, OR 97403, USA} \\\vskip 0.5em
\small \textit{${}^2$Center for Theoretical Physics -- a Leinweber Institute,}\\
\textit{Massachusetts Institute of Technology, Cambridge, Massachusetts 02139, USA}\\ \vskip 0.5em
\textit{${}^3$Department of Physics, Harvard University, Cambridge, Massachusetts 02138, USA}\\ \vskip 0.5em
\textit{${}^4$TRIUMF, 4004 Wesbrook Mall, Vancouver, BC V6T 2A3, Canada}\\ \vskip 0.5em
\textit{${}^5$Department of Physics and Astronomy, University of California, Riverside, CA 92521, USA}
}
}

\AtBeginShipoutNext{\AtBeginShipoutUpperLeft{%
  \put(\dimexpr\paperwidth-1cm\relax,-1.5cm){\makebox[0pt][r]{MIT-CTP/5873}}%
}}

\begin{document}

\maketitle
\begin{abstract}
We introduce a new mechanism for the simultaneous generation of baryon and dark matter asymmetries through 
ultraviolet-dominated freeze-in scatterings. The mechanism relies on heavy Majorana neutrinos that connect the visible Standard Model sector to a dark sector through the neutrino portal. 
Following reheating of the visible sector to a temperature well below the heavy neutrino masses, we show that 2-to-2 scattering processes can populate the dark sector and generate both baryon and dark matter asymmetries. 
In some parameter regions, the dominant source of baryon asymmetry can be charge transfer from the dark sector, a process we call dark wash-in.
We also demonstrate that annihilation of the dark matter to massless states within the dark sector can deplete the symmetric population without destroying the net baryon charge to leave only an asymmetric dark matter abundance today. Depending on the specific model parameters, the observed baryon and dark matter abundances can be attained with heavy neutrino masses $M_N \gtrsim 10^{10} \gev$, and dark matter masses in the range $0.1 \gev \lesssim m_\chi \lesssim 10^3 \gev$ if the dark matter relic abundance is mainly asymmetric and even lower masses if it is symmetric. 
\end{abstract}

\newpage 

\begin{spacing}{1.2}


\tableofcontents

\newpage

\section{Introduction}
\label{sec:intro}

The origin of dark matter~(DM) and the baryon asymmetry are two unresolved puzzles in particle physics and cosmology that point towards new physics beyond the Standard Model~(SM). Each of these open problems has motivated numerous theoretical and experimental studies, see Refs.~\cite{Jungman:1995df,Bertone:2004pz,Roszkowski:2017nbc,Arbey:2021gdg,Cirelli:2024ssz,Bozorgnia:2024pwk} and Refs.~\cite{Riotto:1998bt,Cline:2006ts,Davidson:2008bu,Canetti:2012zc,Morrissey:2012db,Elor:2022hpa,Barrow:2022gsu}, respectively, for recent reviews on DM and baryogenesis. 

Intriguingly, the cosmological energy densities of dark matter and baryons are very similar, $\rho_{\rm DM} \simeq 5\rho_B$~\cite{Planck:2018vyg,ACT:2025fju}. This similarity is highly surprising if the cosmic origins of the two are unrelated, and is suggestive of a common source for both. 
Crucially, such a cogenesis requires non-gravitational interactions between the SM and dark matter.
As this is the only hint of such interactions,
theories that dynamically explain both the SM baryon asymmetry and DM abundance are important targets for future experimental searches.

Numerous pathways exist for producing the DM abundance.
Here, we study freeze-in~\cite{Dodelson:1993je,McDonald:1993ex,Hall:2009bx}, whereby the DM abundance is initially negligible following reheating and is produced through portal
interactions with the SM bath that are too feeble to bring DM into full equilibrium. 
While many freeze-in models focus on renormalizable portal interactions, known as infrared (IR) freeze-in~\cite{Hall:2009bx}, in this work we focus on a non-renormalizable portal, \textit{i.e.} ultraviolet (UV) freeze-in~\cite{Chung:1998rq,Giudice:2000ex,Elahi:2014fsa}. 
In this type of model, the dark sector is primarily populated near the reheat temperature $\Trh$.

A more limited number of mechanisms have been found to explain
the observed baryon asymmetry, which is only possible if they break
lepton or baryon number, violate C and CP, and involve out-of-equilibrium processes~\cite{Sakharov:1967dj}.
For example, in leptogenesis heavy neutrinos generate a lepton asymmetry, which is then converted to a baryon asymmetry by sphaleron transitions~\cite{Fukugita:1986hr,Luty:1992un,Giudice:2003jh,Buchmuller:2005eh,Chen:2007fv,Fong:2012buy}. 
Generically, leptogenesis models rely primarily on the \textit{decay} of
long-lived heavy neutrinos to furnish the departure from equilibrium~\cite{Pilaftsis:1997jf,Pilaftsis:2003gt}, although see Ref.~\cite{Bento:2001rc,Blazek:2024efd} for approaches based on scattering.

Connecting the DM abundance to the baryon asymmetry typically requires further structure. A promising framework to do so is asymmetric dark matter~(ADM)~\cite{Nussinov:1985xr,Kaplan:1991ah, Kaplan:2009ag,Petraki:2013wwa,Zurek:2013wia}.
In the ADM paradigm, the DM abundance is set by an asymmetry between the DM particle and its antiparticle. This allows a connection between the DM and baryon densities when the asymmetry in one sector is related to that in the other, either from the cogenesis of both asymmetries in a single step~\cite{Davoudiasl:2010am, Davoudiasl:2012uw,Falkowski:2011xh,Bell:2011tn,Cheung:2011if, March-Russell:2011ang,Barman:2021tgt,Bose:2024bnp} or through the creation of an asymmetry in one sector that is then shared with the other~\cite{Shelton:2010ta,Haba:2010bm,
Buckley:2010ui,Blennow:2010qp,Hall:2019ank,Hall:2021zsk,Datta:2021elq,Bhattacharya:2021jli}.
Asymmetric dark matter scenarios can be contrasted with models of baryogenesis via weakly interacting massive particles~(WIMPs)~\cite{Cui:2011ab, Cui:2012jh, McDonald:2011zza, Davidson:2012fn, Cui:2015eba, Chu:2021qwk,Mahanta:2022gsi,Heisig:2024mwr} where the baryon or lepton asymmetry is generated during symmetric WIMP freeze-out.\footnote{There have been some attempts to relate the WIMPy baryogenesis paradigm to ADM. 
For example, in Ref.~\cite{Cui:2020dly} the DM abundance is set ultimately by a combination of freeze-out and an asymmetry.}

In this work, we present a new mechanism for the creation of baryon and dark matter asymmetries through freeze-in \textit{scattering} processes. 
The model includes a neutrino portal~\cite{Pospelov:2007mp} that connects the visible SM sector to a dark sector containing SM-singlet fermions and a scalar. 
The heavy neutrinos in our model are assumed to be heavier than the SM bath reheating temperature $\Trh$. Transfer reactions between the two sectors mediated by the heavy neutrinos populate the dark sector through UV freeze-in and generate charge asymmetries in both sectors through $2\to 2$ scatterings. The asymmetry in the visible sector is partially reprocessed into the observed baryon asymmetry. The related asymmetry in the dark sector impacts the final dark matter relic density.

Freeze-in is an essential component of scattering cogenesis, as it gives rise to the required out-of-equilibrium Sakharov condition~\cite{Toussaint:1978br,Weinberg:1979bt,Nanopoulos:1979gx,Kolb:1979qa,Barr:1979wb}. However, asymmetry creation by freeze-in is strongly constrained by CPT and unitarity~\cite{Hook:2011tk}.
In particular, in scenarios with an exactly conserved global charge, the leading charge asymmetry transfer between sectors vanishes~\cite{Hook:2011tk,Unwin:2014poa}.
As a result, the parameter space that generates sufficient SM baryon asymmetry (at next-to-leading order in the feeble portal coupling) also generates a high DM symmetric number density (at leading-order), which is typically ruled out by a combination of the overclosure constraint~\cite{Planck:2018vyg} and other astrophysical observations~\cite{Sigurdson:2009uz,Das:2010ts}.

Two main aspects of our model allow us to overcome the challenges outlined in Ref.~\cite{Hook:2011tk}. 
First, our model includes a natural explicit breaking of the extended lepton number between the two sectors. This allows for an enhanced charge creation relative to the class of scenarios treated in Ref.~\cite{Hook:2011tk}. Second, we postulate that our 
dark matter candidate is in contact with a bath of light secluded dark particles 
that acts as a \textit{dark sink}~\cite{Blennow:2012de,Bhattiprolu:2023akk, Bhattiprolu:2024dmh}; eventually the symmetric abundance of DM freezes out into this dark sink, reducing its population enormously and
leaving only an asymmetric DM abundance today. 
The dark sink redshifts like radiation and is not currently constrained by cosmological tests of new relativistic degrees of freedom.
Thanks to these two features ((i) the breaking of the extended lepton number and (ii) the dark sink), our viable dark matter mass range extends beyond the standard ADM models to DM masses $1 \keV \lesssim m_\chi \lesssim 10^3 \gev$, while its abundance can be dominated by the asymmetry for $m_\chi \gtrsim 0.1 \gev$.\footnote{See Refs.~\cite{Falkowski:2011xh, Hall:2021zsk} for other ADM models viable over a wide range of DM masses.}  

The higher end of this mass range is achieved in the strong dark wash-out limit.
Wash-out terms play a key role in our setup. 
Existing constraints on the active neutrino masses push us 
to a parameter region where the dark lepton number source terms are more efficient than those in the visible sector. 
Some of the dark asymmetry from these sources is transferred to the visible sector through wash-out processes. In leptogenesis, the concept of wash-out terms reprocessing a ``primordial charge asymmetry" into a SM $B-L$ asymmetry is known as \textit{wash-in}~\cite{Domcke:2020quw, Domcke:2022kfs, Mojahed:2025vgf}. In our case, the primordial asymmetry lies in the dark sector charge density, so we refer to this mechanism as \textit{dark wash-in}.

Our model is a complete UV freeze-in cogenesis setup to produce asymmetric dark matter. The underlying asymmetry creation mechanism is closely related to that of Ref.~\cite{Bento:2001rc}, while other more recent studies have followed somewhat different approaches~\cite{Baldes:2015lka,Goudelis:2021qla,Goudelis:2022bls,Blazek:2025wmc,Shuve:2020evk,Flood:2021qhq,Berman:2022oht}, many using new elements such as the ARS mechanism~\cite{Akhmedov:1998qx} or resonant leptogenesis~\cite{Pilaftsis:2003gt}. In Ref.~\cite{Baldes:2015lka} the DM abundance can be asymmetric, while in Refs.~\cite{Goudelis:2021qla,Goudelis:2022bls,Blazek:2025wmc,Shuve:2020evk,Flood:2021qhq,Berman:2022oht} it is dominantly symmetric.
Our setup also shares different moving pieces with existing models of baryogenesis that do not attempt to explain the DM abundance~\cite{Baldes:2014gca,Baldes:2014rda,Blazek:2024efd}.

The rest of the paper is organized as follows. 
In Sec.~\ref{sec:model} we introduce our model and give an overview of its different moving pieces, including its cosmological evolution and the relevant processes for generating an asymmetry. 
In Sec.~\ref{sec:cosmo} we provide more details on the cosmic evolution of our setup and populating the dark sector. 
Details of the asymmetry generation are provided in Sec.~\ref{sec:BE}, where we include the final form of the Boltzmann equations and discuss the significance of the wash-out terms. 
Finally, in Sec.~\ref{sec:sink} we discuss further dynamics in the dark sector and calculate the viable DM mass range. 
We conclude in Sec.~\ref{sec:conclusion}. 
In App.~\ref{appx:collterm} we provide further details on the collision terms appearing in the Boltzmann equations. 
Further details on the calculation of asymmetric matrix elements can be found in App.~\ref{appx:Mcalc}. 
In App.~\ref{appx:fullBE} we review the steps taken to simplify the Boltzmann equations. 
{
In App.~\ref{appx:reheat} we compute energy transfer from the visible sector to the dark sector during non-instantaneous reheating after inflation.}

\section{Particles, Interactions, and Processes}
\label{sec:model}

In this section we introduce a simple model that can realize asymmetric DM and leptogenesis through scattering processes during UV freeze-in.
It consists of the SM together with heavy Majorana neutrinos that connect to a dark sector containing a Dirac fermion, a complex scalar, and a massless Weyl fermion.
We also give an overview of the roles of these states in the cosmological production of the baryon asymmetry and dark matter density, and we calculate the relevant interaction cross sections for the corresponding processes. These results will be applied in the following sections to compute the detailed cosmological evolution of this theory.

\subsection{Overview of the Model}
\label{subsec:L}

In addition to the SM, we consider new Majorana neutrinos $N_a$ together with a dark sector consisting of a Dirac fermion with chiral components $\chi$ and $\chi^c$, a complex scalar $\phi$, and a chiral fermion $\eta$. The interaction Lagrangian for these new fields (in 2-component notation) is
\beq
\lag  = \lag_{SM} + \lag_{asym} + \lag_{sink} \ ,
\eeq
with
\beq
-\lag_{asym}   \ \supset \    
y_{ia} H\ccdot\ell_i\,N_a + \lambda_{a} \phi\,\chi\,N_a + m_\chi \chi\chi^c+ {m_\phi^2} \phi^\dagger \phi + \frac{1}{2}M_{a} N_a N_a + {\rm h.c.}
\label{eq:L} \ ,
\eeq
and
\beq
-\lag_{sink} = y_\chi\,\phi^*\,\chi\,\eta + {\rm h.c.} ~.
\label{eq:sink}
\eeq
Here, $\ell_i$ denotes the lepton doublets of the SM in the standard lepton flavor-diagonal basis, $H$ is the SM Higgs field with its SU(2)$_L$ indices contracted asymmetrically with $\ell$, and $i$ and $a$ refer to flavors of leptons and heavy neutrinos, respectively. The Yukawa couplings $y_{ia}$ and $\lambda_{a}$ are generic complex matrices. For simplicity, we make the non-essential assumption that the $N_a$ couple equally to all SM leptons, $y_{ia} = y_a$, $i=1,2,3$. We also consider two generations of $N_a$ in the spectrum ($a=1,2$) with $M_1 < M_2$.

\begin{table}
\centering
\begin{tabular}{c c c c c }
\toprule
• & U(1)$_Y$ & U(1)$_{D}$ & U(1)$_L$ & U(1)$_{L_x}$  \\ 
\midrule
$H$ & $+1/2$ & 0 & \phantom{+}0 & 0 \\ 
$\ell$ & $-1/2$ & 0 & $+1$ & 0  \\ 
$\phi$ & 0 & $-1/2$ & \phantom{+}0 & $+1/2$ \\ 
$\chi$ & 0 & $+1/2$ & \phantom{+}0 & $+1/2$ \\ 
$\chi^c$ & 0 & $-1/2$ & \phantom{+}0 & $-1/2$ \\ 
$N$ & 0 & 0 & $-1$ & 0 \\
$\eta$ & 0 & $-1$\phantom{/2} & \phantom{+}0 & 0 \\
\bottomrule
\end{tabular} 
\caption{The field content and (approximate) symmetries of the Lagrangian from Eq.~\eqref{eq:L} in our UV freeze-in DM setup. We define the symmetry U(1)$_D$, similar to SM hypercharge, in the dark sector. The extended lepton number, U(1)$_{L+L_x}$, is explicitly broken by the Majorana mass of $N$ particles. These particles act as a (neutrino) portal between SM and the dark sector. The $\chi$ field is the lightest particle with fractional U(1)$_D$ charge; hence it is stable and acts as the DM in this scenario. }
\label{tab:charges}
\end{table}

In the limit of vanishing Majorana masses $M_a\to 0$, the Lagrangian has three independent Abelian symmetries. They can be chosen to be: (i) the usual SM hypercharge $Y$, (ii) a dark hypercharge ${D}$, and (iii) an extended lepton number $L+L_x$, where $L_x$ is an effective lepton number in the dark sector.  The charges of the various fields under these symmetries are given in Tab.~\ref{tab:charges}. Adding Majorana masses for $N_a$ breaks $L+L_x$ explicitly and softly. Taking the opposite limit of $M_a \to \infty$, $L$ and $L_x$ become approximate independent symmetries in the low-energy effective theory up to explicit breaking by operators suppressed by $M_a$. We note further that the interactions of Eq.~\eqref{eq:L} are the most general ones possible at the renormalizable level up to {a $|\phi|^4$ quartic self-coupling and a $|\phi|^2|H|^2$ Higgs portal coupling. A quartic self-coupling for $\phi$ would not significantly affect the dynamics of the mechanism. However, a Higgs portal coupling connecting the dark and visible sectors could lead to the thermalization of the sectors with each other at late times. We assume this interaction to be negligibly small for our analysis, but we comment on the impact of larger values later on.}

Creation of baryon or lepton asymmetries requires new sources of CP violation. Here, this can arise if some of the couplings in Eq.~\eqref{eq:L} are complex. 
Without loss of generality, we take the $M_a$ to be real, positive, and diagonal. 
After field redefinitions that preserve this convention, there remain $(n_N-1)$ physical phases in each of $y$ and $\lambda$, and thus our choice of $n_N=2$ is the minimal possibility consistent with new CP violation. This remaining freedom can be used to make $y_2$ and $\lambda_2$ real and positive, such that the two new physical phases come solely from $y_1$ and $\lambda_1$.

In the dark sector, the lighter of $\chi$ and $\phi$ is stable on account of its smallest fractional U(1)$_D$ charge. To be concrete, we take $m_\chi < m_\phi$, meaning that $\chi$ (combined with its vector-like partner $\chi^c$) is a stable Dirac fermion dark matter candidate. The massless Weyl fermion~$\eta$ acts as a dark sink~\cite{Bhattiprolu:2023akk,Bhattiprolu:2024dmh}, providing an annihilation channel for $\chi$ as well as a decay channel for $\phi$ through $\phi \to \chi \eta$. This interaction also enables rapid self-thermalization within the dark sector.\footnote{If the U(1)$_D$ dark hypercharge were a global symmetry, we would expect quantum gravity effects to break it~\cite{Banks:2010zn}, which could destabilize $\chi$. This could be avoided by weakly gauging the U(1)$_D$, which would also require introducing a vector-like partner $\eta^c$ for anomaly cancellation. Including such a state would not significantly alter our conclusions provided the $\eta$-$\eta^c$ Dirac mass is very small.}

This scenario is an example of a neutrino portal dark sector. Such portals offer the possibility of explaining (at least part of) the SM neutrino masses through the first term of Eq.~\eqref{eq:L}. For the universal flavor structure $y_{ia}=y_a$, the neutrino mass matrix has two heavy mostly-singlet states with masses near $M_1$ and $M_2$, two massless active neutrinos, and one massive active neutrino with Majorana mass in the seesaw limit of $M_a\gg m_Z$, where $m_Z$ represents the weak scale, equal to
\beq
\label{eq:seesaw}
\Delta m_\nu \ = \ \frac{3}{2}\left|\frac{y_1^2v^2}{M_1}+\frac{y_2^2v^2}{M_2}\right| \ ,
\eeq
where $v \simeq 246 \gev$ is the Higgs vacuum expectation value. This massive active neutrino is an equal linear combination of all three flavor eigenstates.
In this work we do not attempt to explain the full pattern of observed neutrino masses and mixings. However, we do check that the contribution above to the sum of the active neutrino masses is not too large by imposing the constraint
\beq
\Delta m_\nu < 0.07 \eV \ ,
\label{eq:nubound}
\eeq
which is motivated by the cosmological bound $\sum m_{\nu} < 0.072 \eV$~\cite{DESI:2024mwx}.

\subsection{Cosmological Evolution and Roles}

Having introduced the fields and interactions above, our next step is to give an overview of the cosmological timeline over which these fields can generate the baryon asymmetry and dark matter density. In the visible sector, some of the lepton asymmetry is converted to a baryon asymmetry by electroweak sphaleron transitions. In the dark sector, the heavy $\phi$ scalars decay to the lighter $\chi$ fermions, which subsequently annihilate efficiently to massless $\eta$ fermions. This leaves a relic density of $\chi$ as dark matter that may be dominated by the dark sector asymmetry. Detailed calculations of the cosmological evolution of all the relevant species densities will be given in upcoming sections.

\begin{figure}[t]
\centering
\includegraphics[width=\textwidth]{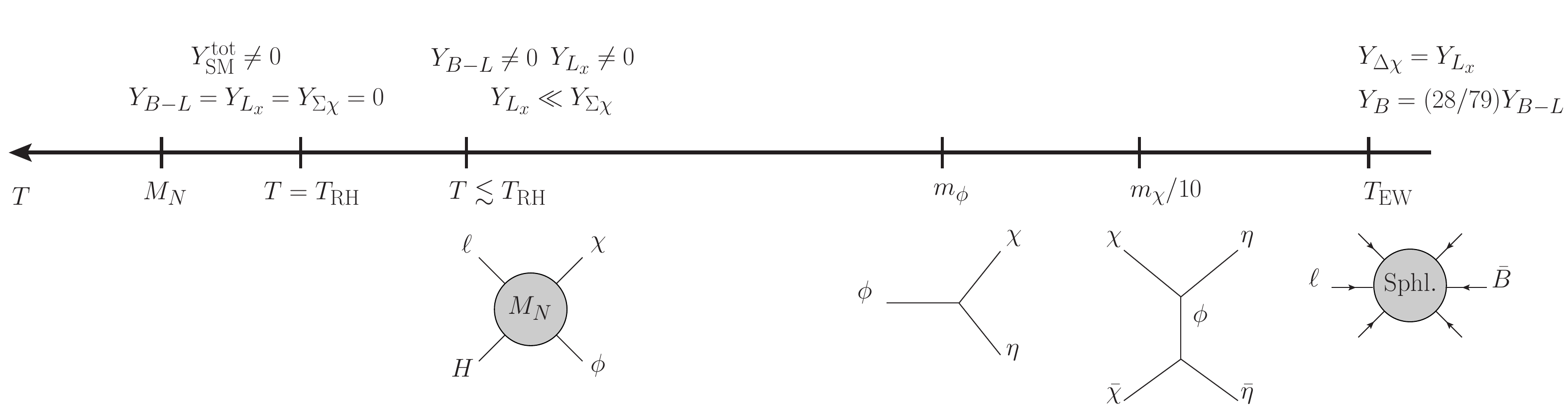}
\caption{Schematic of the cosmological history and the evolution of different abundances $Y_Q$ in our setup. Temperature increases from right to left and time runs from left to right. The universe reheats to temperature $\Trh$ below the heavy neutrino masses and only populates the SM. The dark sector is immediately populated through UV freeze-in. The asymmetries in the two sectors are dominantly generated at this stage as well. At lower temperatures, the $\phi$ field decays to the DM $\chi$, which subsequently freezes-out into the dark bath $\eta$. 
If the frozen-out symmetric abundance of $\chi$ (denoted by $Y_{\Sigma \chi}$) is not too far from the asymmetric yield $Y_{\Delta \chi}$, the freeze-out process produces mostly asymmetric dark matter ($Y_{\Delta \chi} \sim Y_{\Sigma \chi}$).
The generated SM lepton asymmetry is partially converted into a baryon asymmetry by SM sphalerons throughout this timeline; the conversion stops after the electroweak symmetry breaking.
}\label{fig:cartoon}
\end{figure}

The cosmological history in our scenario is illustrated in Fig.~\ref{fig:cartoon}. 
Here $Y_Q$ denotes the yield of a quantum charge $Q$ and $Y_{\Delta\chi,\Sigma\chi} = Y_\chi\mp Y_{\bar{\chi}}$ are the asymmetric and symmetric yields.
Key stages of this evolution are: 
\begin{enumerate}
    \item Primordial inflation reheats the visible sector to a temperature $T=\Trh \ll M_a$ as well as $\Trh \gg m_\phi,\,m_\chi,\,m_Z$.
    We assume further that this reheating  
    populates exclusively the visible sector with no significant densities of $N_a$ or dark states $\chi$, $\phi$, or $\eta$ created at this point. 
    \item Immediately after reheating, scattering processes initiated by $\ell$ and $H$ within the visible sector begin to populate the dark sector. Since these processes are all mediated by the heavy $N_a$ with masses $M_a \gg \Trh$, the direct production of $N_a$ is negligible and transfer to the dark sector is dominated by $2\to 2$ reactions at temperatures close to reheating, corresponding to UV freeze-in production~\cite{Elahi:2014fsa}. As this occurs, scattering within the dark sector through $\eta$ quickly equilibrates this sector to a dark temperature $T_x < T$. 
    \item With C and CP violation together with $L$ number violation from the massive Majorana neutrino masses $M_a$, $2\to 2$ scattering processes between and within the visible and dark sectors generate $B-L$ and $L_x$ charge asymmetries in each sector. Wash-out processes also deplete the asymmetries and transfer them from one sector to the other. Just like the reactions populating the dark sector, these asymmetry transfer processes are dominated by temperatures near reheating and tend to freeze out soon after. Once the transfer reactions freeze out, the two sectors dilute and cool independently with their respective asymmetries remaining fixed.
    \item  As the dark temperature falls below $T_x \sim m_\phi$, the asymmetry in $\phi$ is transferred to the one in $\chi$ by decays $\phi \to \chi \eta$.
    When $T_x$ falls below $m_\chi$, the massive dark fermions undergo freeze-out through ${\chi\bar{\chi}\to \eta \bar{\eta}}$,
    depleting their symmetric density 
    but leaving the 
    asymmetry unchanged. The massless $\eta$ fermion thus serves both to self-thermalize the dark sector and to absorb its symmetric abundance, \textit{i.e.} acting as a dark sink in the sense of Refs.~\cite{Bhattiprolu:2023akk,Bhattiprolu:2024dmh}.
    \item In the visible sector, a fraction of the visible $B-L$ asymmetry is converted into a baryon asymmetry by electroweak sphaleron transitions throughout this timeline~\cite{Khlebnikov:1988sr}. This conversion stops shortly after electroweak symmetry breaking~\cite{DOnofrio:2014rug}.
\end{enumerate}

Our scenario has all the ingredients required to generate the DM abundance and the SM baryon asymmetry. Specifically, the scenario realizes the well-known Sakharov conditions for creating the matter asymmetry~\cite{Sakharov:1967dj,Cline:2006ts}. Violation of lepton number~$L$ is provided by the Majorana masses~$M_a$ of the heavy neutrinos, while electroweak sphaleron transitions affect the combination $B+L$. Breaking of C and CP are provided by the new chiral interactions together with the weak phases introduced by the new couplings of Eq.~\eqref{eq:L}. And finally, a departure from equilibrium arises if the visible and dark sectors do not fully equilibrate with each other. 
In the sections to come, we will show explicitly how these ingredients can combine to produce asymmetries in visible baryon charge and dark matter number.

\subsection{Overview of Transfer Reactions}
\label{subsec:overview_reactions}

The first stage of cosmological evolution in our scenario after primordial reheating to a visible temperature $\Trh$ is dictated
by \emph{transfer reactions}. We define these to be the reactions mediated by the heavy Majorana neutrinos $N_a$. At leading order in the relevant couplings, transfer reactions share particle number and energy between the visible and dark sectors. Going beyond leading order, a subset of these reactions can also generate asymmetries within and between the two sectors. 
We outline our calculation of transfer reaction rates in this subsection. 

It is useful to define some quantities of interest and introduce a simplified notation. We write the matrix element for a process going from the initial state $\alpha$ to the final state $\beta$ as 
\begin{equation}
    \M(\alpha\rightarrow\beta) \equiv \M^{\alpha}_{\beta} \ .
    \label{eq:defMnotation}
\end{equation}
The matrix elements for a process and its charge conjugate can be merged to define symmetric and asymmetric squared matrix elements:
\begin{align}
     |\Sigma\M^\alpha_\beta|^2 &\equiv |\M^\alpha_\beta|^2 + |\M^{\alpha^\dagger}_{\beta^\dagger}|^2  \ ,
    \label{eq:defSigmas}
    \\
    |\Delta \M^\alpha_\beta|^2 &\equiv |\M^\alpha_\beta|^2 - |\M^{\alpha^\dagger}_{\beta^\dagger}|^2 \ .
    \label{eq:defDeltaM}
\end{align}
In these and other squared matrix elements, we implicitly sum over all initial and final internal degrees of freedom. We also write
\begin{align}
    n_{\Sigma X} &\equiv n_X + n_{X^\dagger} \ ,
    \label{eq:sigmandef}
    \\
    n_{\Delta X} &\equiv n_X - n_{X^\dagger} \ ,
    \label{eq:deltandef}
\end{align}
where $X$ denotes any particle in the spectrum and $X^\dagger$ its antiparticle. The evolution of these symmetric and asymmetric abundances depends on the symmetric and asymmetric matrix elements. 

\begin{table}[t]
    \centering
    \begin{tabular}{r c l c c c c}
    \toprule
        \multicolumn{3}{c}{Reaction} & $\phantom{AAA}\Delta L$\phantom{AAA} &\phantom{AAA}$\Delta L_x$\phantom{AAA} & \phantom{A}$\Delta (L+L_x)$\phantom{A} & Asymmetry?  \\
        \midrule 
    $\ell_i H$ & $\to$ & $\ell_j^\dagger H^\dagger$ & $-2$ & \phantom{+}0 & $-2$ & Y\\[0.1cm] 
    $\ell_i H$ & $\to$ & $\chi^\dagger\phi^\dagger$ & $-1$ & $-1$ & $-2$ & Y\\[0.1cm] 
    $\ell_i H$ & $\to$ & $\chi\phi$ & $-1$ & +1 & \phantom{+}0 & Y\\[0.1cm] 
    $\chi\phi$ & $\to$ & $\chi^\dagger \phi^{\dagger}$ & $\phantom{+}$0 & $-2$ & $-2$ & Y\\[0.1cm]
    \midrule
    $\ell_i \ell_j$ & $\to$ & $H^\dagger H^\dagger$ & $-2$ & \phantom{+}0 & $-2$ & N\\[0.1cm] 
    $\ell_i\chi$ & $\to$ & $H^\dagger \phi^\dagger$ & $-1$ & $-1$ & $-2$ & N\\[0.1cm] 
    $\ell_i\chi^\dagger$ & $\to$ & $H^\dagger \phi$ & $-1$ & $+1$ & \phantom{+}0 & N\\[0.1cm] 
    $\ell_i \phi$ & $\to$ & $H^\dagger \chi^\dagger$ & $-1$ & $-1$ & $-2$ & N\\[0.1cm] 
    $\ell_i \phi^\dagger$ & $\to$ & $H^\dagger \chi$ & $-1$ & $+1$ & \phantom{+}0 & N\\[0.1cm] 
    $\chi \chi$ & $\to$ & $\phi^\dagger \phi^\dagger$ & \phantom{+}0 & $-2$ & $-2$ & N\\[0.1cm]
    \bottomrule
    \end{tabular}
    \caption{List of all relevant $2\to 2$ transfer reactions in our scenario, up to time reversal and charge conjugation, as well as the changes in charges they induce.  They are separated according to whether~(Y) or not~(N) they give rise to non-zero asymmetric matrix elements at leading non-trivial order. 
    }
    \label{tab:transfer}
\end{table}

The transfer reactions relevant to our scenario are all $2\to 2$ processes and listed in Tab.~\ref{tab:transfer}. We divide these processes into two broad categories. Those in the upper section of the table contribute to generating particle asymmetries. In contrast, reactions in the lower section of the table do not create asymmetries at leading non-trivial order, but they can have an important impact on particle asymmetries by washing out asymmetries that have already been created. Further details on asymmetry creation and wash-out will be discussed below.

Since transfer reactions are mediated by the exchange of heavy Majorana neutrinos and we focus on temperatures well below their masses, $T\ll M_a$, it is a good approximation to expand their corresponding matrix elements in powers of $1/M_a$. The expansion begins at the linear order, and each term is accompanied by one or more positive powers of energy $E$. To compute collision terms for the evolution of number or energy densities, these matrix elements are then integrated over the final and initial state phase spaces weighted by the appropriate quantum distribution functions and statistics factors. The appearance of positive powers of energies implies that the support for these integrals is pushed to regions with ${E/T \gtrsim \mathcal{O}(\text{few})}$ where the quantum distributions are numerically close to Maxwell-Boltzmann~(MB)~\cite{Bringmann:2021sth}. Motivated by this, we make the simplifying approximation of evaluating all transfer reaction collision terms using the MB approximation in this work.

Within the MB approximation, it is convenient to define for each process of the form $1+2\to 3+4$ a Lorentz invariant \emph{scattering kernel}:
\beq
\mathcal{W}^{12}_{34}(s)
\ \equiv \ {S}\!\int\!{d\Pi_3}\!\int\!{d\Pi_4}\;(2\pi)^4\delta^{(4)}(p_1+p_2-p_3-p_4)|\,\mathcal{M}^{12}_{34}|^2 \ ,
\label{eq:wscatt}
\eeq
where $s=(p_1+p_2)^2$ is the Mandelstam variable, $d\Pi_i = d^3p_i/[(2\pi)^32E_i]$ are phase space measures, ${S}$ is the usual symmetry factor for identical initial or final states, and the squared matrix element is implicitly summed over all initial and final internal degrees of freedom. A scattering kernel is related to the scattering cross section of the corresponding process according to $\mathcal{W} = 4\,E_1E_2\,g_1g_2\,\sigma v$, where $g_i$ is the number of internal degrees of freedom of particle $i$ and $v$ is the relative (M\o ller) velocity of the incoming particles~\cite{Gondolo:1990dk,Edsjo:1997bg}.

Given a scattering kernel for a reaction, its contribution to the collision term for the evolution of an observable $\mathcal{O}$ that is independent of the final-state phase space in the MB approximation is  
\beq
[{{C}}_{\mathcal{O}}]^{12}_{34}
= \int\!d\Pi_1\!\int\!d\Pi_2\,\mathcal{W}^{12}_{34}(s)\;f_1f_2\;\delta\mathcal{O} \ ,
\label{eq:collision}
\eeq
where $f_i$ are the MB distribution functions for the initial states and $\delta\mathcal{O}$ is the change in the observable in the reaction. In particular, for number density of a species, $\delta\mathcal{O}$ is the net change in the species particle number per reaction. The full collision term is the sum of contributions from all relevant reactions. Further details on evaluating collision terms are collected in App.~\ref{appx:collterm}.

The expansion of matrix elements in powers of $1/M_a$ discussed above carries through to the scattering kernels for symmetric and asymmetric squared matrix elements. By Lorentz invariance, the positive powers of energy appearing in matrix elements must end up as positive powers of the Mandelstam variable $s$. This allows us to write
\beq
\Sigma\mathcal{W}^{\alpha}_\beta(s)  &\equiv &
\mathcal{W}^{\alpha}_\beta(s) + \mathcal{W}^{\alpha^\dagger}_{\beta^\dagger}(s) \ \ \equiv \ \
\losigma^\alpha_\beta \times s \hspace{2mm} + \ \nlosigma^\alpha_\beta \times s^2 \hspace{2mm}  + \ \mathcal{O}(s^3) \ ,
\label{eq:defexpansionssigma}
\\
\Delta\mathcal{W}^{\alpha}_\beta(s)  &\equiv &
\mathcal{W}^{\alpha}_\beta(s) - \mathcal{W}^{\alpha^\dagger}_{\beta^\dagger}(s) \ \ \equiv \ \
\lodelta^\alpha_\beta\times s  \ + \ \nlodelta^\alpha_\beta \times s^2 \ + \ \mathcal{O}(s^3)    \ .
\label{eq:defexpansionsdelta}   
\eeq
The coefficients $[\mathcal{S}_{i}]^\alpha_\beta$ and $[\mathcal{D}_i]^\alpha_\beta$ are independent of $s$. As demonstrated in Ref.~\cite{Elahi:2014fsa} and generalized in App.~\ref{appx:collterm}, expanding this way allows for a simple connection between these coefficients and the collision terms that control the cosmological evolution of number and energy densities.

Turning now to explicit calculations in our theory, we collect in  Tab.~\ref{tab:Mvalues} the tree-level symmetric coefficients $\losigma$ and $\nlosigma$ for all relevant transfer reactions.\footnote{The expression for $\ell_i\ell_j\to H^\dagger H^\dagger$ includes a symmetry factor for the initial state appropriate for $i=j$. Later on we will sum these terms freely over all lepton flavors $i$ and $j$. This sum double counts when $i\neq j$ but is compensated for by the extra factor of $1/2$ included in the table.} This is a complete set of symmetric coefficients since by definition the symmetric matrix element for the process $\alpha\to \beta$ is equal to that for $\alpha^{\dagger}\to \beta^{\dagger}$, while CPT implies equality between those for $\alpha\to \beta$ and $\beta^\dagger \to \alpha^\dagger$. Some of the $\losigma$ coefficients in Tab.~\ref{tab:Mvalues} vanish, which can be understood in terms of the charge $L+L_x$. This charge is only broken by the Majorana masses $M_a$, which act as a spurion. Complexifying this mass and counting its insertions show that $\losigma = 0$ for reactions that preserve $L+L_x$.
In contrast to the symmetric matrix elements, all asymmetric matrix elements vanish at tree-level and only emerge at loop order. These will be investigated in detail below.

\begin{table}[ttt]
    \centering
    \begin{tabular}{r c l  c  c}
    \toprule
        \multicolumn{3}{c}{Process}   & $4\pi\losigma$ & $4\pi\nlosigma$  \\
        \midrule
        $\ell_i H$ & $\to$ & $\ell_j^\dagger H^\dagger $  &  \phantom{AA}$6 \abs{\sum_a y_{ia} M_a^{-1} y_{ja}}^2$\phantom{AA} & $4 \Re \left[ \sum_a y_{ia} M_a^{-3} y_{ja} \sum_b y_{ib}^\ast M_b^{-1} y_{jb}^\ast \right]$ \\[0.2cm] 
        
        $\ell_i H$ & $\to$ & $\chi^\dagger \phi^\dagger$  & $\abs{\sum_a \lambda_{a} M_a^{-1} y_{ia}}^2$ & $2 \Re \left[ \sum_a \lambda_{a} M_a^{-3} y_{ia} \sum_b \lambda_{b}^\ast M_b^{-1} y_{ib}^\ast \right]$ \\[0.2cm] 
        
        $\ell_i H$ & $\to$ & $\chi \phi$   & $0$ & $\abs{\sum_a \lambda_a^\ast M_a^{-2} y_{ia}}^2$ \\[0.2cm] 
        
        $\chi \phi$ & $\to$ & $ \chi^\dagger \phi^\dagger  $   & $2 \abs{\sum_a \lambda_a^2 M_a^{-1}}^2$ & $(4/3) \Re \left[ \sum_a \lambda_a^2 M_a^{-3} \sum_b \lambda_b^{\ast 2} M_b^{-1} \right]$ \\[0.2cm]

        $\ell_i\ell_j$ & $ \to$ & $ H^\dagger H^\dagger $    &$3|\sum_ay_{ia}M_a^{-1}y_{ja}|^2$ &  $3\Re \left[ \sum_a y_{ia} M_a^{-3} y_{ja} \sum_b y_{ib}^\ast M_b^{-1} y_{jb}^\ast \right]$\\[0.2cm]

        $\ell_i\chi$ & $ \to$ & $ H^\dagger \phi^\dagger $   & $2 \abs{\sum_a \lambda_a M_a^{-1} y_{ia}}^2$ & $-2 \Re \left[ \sum_a \lambda_a M_a^{-3} y_{ia} \sum_b \lambda_b^\ast M_b^{-1} y_{ib}^\ast \right]$ \\[0.2cm] 

        $\ell_i\chi^\dagger $ & $\to$ & $ H^\dagger \phi $    & $0$ & $(1/3) \abs{\sum_a \lambda_a^\ast M_a^{-2} y_{ia}}^2$ \\[0.2cm] 

        $\ell_i\phi $ & $\to$ & $ H^\dagger \chi^\dagger $    & $\abs{\sum_a \lambda_a M_a^{-1} y_{ia}}^2$ & \phantom{A}$(-2/3) \Re \left[ \sum_a \lambda_a M_a^{-3} y_{ia} \sum_b \lambda_b^\ast M_b^{-1} y_{ib}^\ast \right]$\phantom{A} \\[0.2cm] 

        $\ell_i\phi^\dagger $ & $\to$ & $ H^\dagger \chi $    & $0$ & $\abs{\sum_a \lambda_a^\ast M_a^{-2} y_{ia}}^2$ \\[0.2cm] 

        $\chi\chi$ & $ \to$ & $ \phi^\dagger \phi^\dagger $    & $ \abs{\sum_a \lambda_a^2 M_a^{-1}}^2$ & $- \Re \left[ \sum_a \lambda_a^2 M_a^{-3} \sum_b \lambda_b^{\ast 2} M_b^{-1} \right]$ \\[0.2cm]   
        
        \bottomrule
    \end{tabular}
    \caption{Coefficients $\losigma$ and $\nlosigma$ for symmetric scattering kernels defined in Eq.~\eqref{eq:defexpansionssigma} for all relevant transfer reactions in our scenario. They are the first two terms in an expansion in powers of the Mandelstam variable $s$.
    }
    \label{tab:Mvalues}
\end{table}

\subsection{Overview of Asymmetry Generation}
\label{subsec:overview_asymm}

A subset of the transfer reactions listed in Tab.~\ref{tab:transfer} 
have a non-zero asymmetric squared matrix element $|\Delta\mathcal{M}^{\alpha}_{\beta}|^2$, a necessary ingredient for sourcing net charges in the early universe. 
In this section we identify the dominant sources of such asymmetries and compute their asymmetric matrix elements at leading non-trivial order.

Physical CP violation requires interference involving two types of phases:
\begin{itemize}
    \item \textit{Weak phase:} a phase that changes sign between a process and its conjugate. These phases are provided by the couplings in the Lagrangian~\cite{Weinberg:1979bt,Kolb:1979qa}.
    \item \textit{Strong phase:} a phase that {does not} change sign between a process and its conjugate. These arise in the present context from intermediate virtual particles going on shell in loops~\cite{Cutkosky:1960sp}.
\end{itemize}
To demonstrate this explicitly, let us decompose the matrix element for a process $\alpha \rightarrow \beta$ in terms of independent Feynman diagrams as
\begin{equation}
    \M^{\alpha}_\beta = \sum_n\sum_{i_n} c_{i_n} P_{i_n} \ ,
    \qquad
    \M^{{\alpha}^{\dagger}}_{\beta^{\dagger}} \equiv \sum_n\sum_{i_n} c_{i_n}^* P_{i_n} \ ,
    \label{eq:PCnotation}
\end{equation}
where $n\geq 0$ counts the loop order, $i_n$ runs over all contributions at loop order $n$, $c_{i_n}$ contains all the couplings appearing in the corresponding diagram and is conjugated in the conjugate process, and $P_{i_n}$ denotes the rest of the amplitude and is identical for a process and its conjugate. It follows that the asymmetric squared matrix element of Eq.~\eqref{eq:defDeltaM} is given by 
\begin{equation}
    |\Delta \M^\alpha_\beta|^2 = 
    -4\sum_{i_0,i_1}\Im(P_{i_0}^*P_{i_1})\Im(c_{i_0}^*c_{i_1})
    + \ldots
    \ , \label{eq:MinPcnotation}
\end{equation}
where the omitted terms are of higher order in the loop expansion. This expression shows why a non-zero asymmetric squared matrix element is contingent on having both \textit{weak phases}~(from $c_i$) as well as \textit{strong phases}~(from $P_j$). In our scenario, weak phases come from the non-trivial phases in the Yukawa couplings $y_{ia}$ and $\lambda_a$. Strong phases arise from the imaginary parts of loop amplitudes generated by intermediate particles going on shell. 

\begin{figure}[ttt]
    \centering
    \includegraphics[trim={0 0 0cm 0cm}, clip, width=0.8\linewidth]{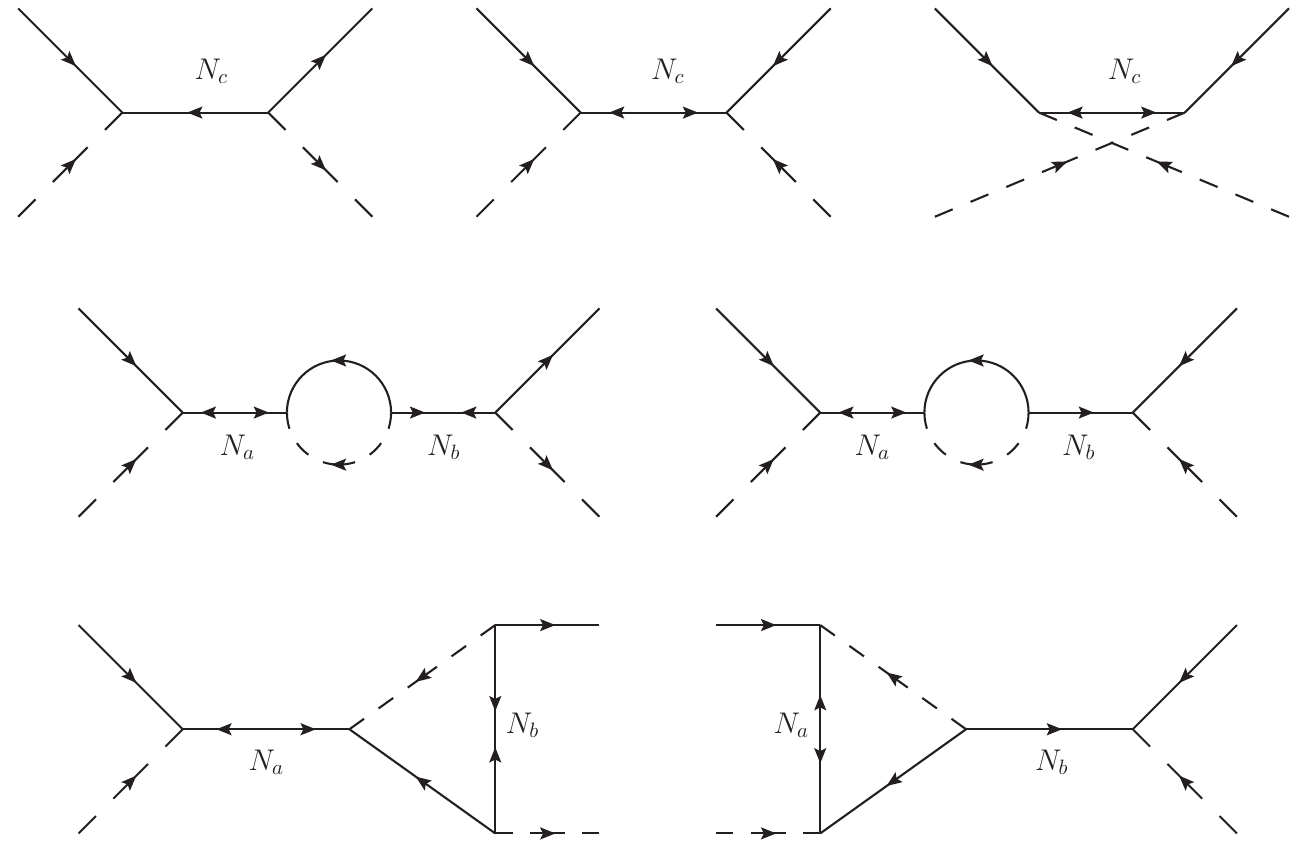}
    \caption{Diagrams contributing to asymmetric squared matrix elements at leading non-trivial order. The tree diagrams in the top left (middle and right) apply to reactions that conserve (violate) $L+L_x$ such as $\ell_i H\to \chi \phi$ ($\ell_i H\to \chi^\dagger \phi^\dagger$) and have an even (odd) number of Majorana neutrino mass insertions. 
    The $u$-channel tree-level diagram only arises for $\ell_i H\to \ell_j^\dagger H^\dagger$, $\chi\phi\to\chi^\dagger\phi^\dagger$, and their conjugates.
    The loop diagrams in the left~(right) column apply to processes that conserve~(violate) $L+L_x$ and involve two~(one) Majorana mass insertion(s).}
    \label{fig:feynman_asym}
\end{figure}

Among the four processes listed in Tab.~\ref{tab:transfer}, the asymmetries in $\ell_i H\to \ell_j^\dagger H^\dagger$ and ${\chi \phi \to \chi^\dagger\phi^\dagger}$ are nearly identical in terms of the amplitude topologies by which they are generated. Asymmetries in both processes arise from the interference between tree-level contributions from $s$- and $u$-channel amplitudes with one-loop diagrams with $s$-channel massive neutrinos $N_a$ receiving propagator corrections from loops of $\chi\phi$~(for $\ell_i H\to \ell_j^\dagger H^\dagger$) or $\ell_k H$~(for $\chi \phi \to \chi^\dagger \phi^\dagger$) and Majorana mass insertions on one or the other side of the loop.
These topologies are illustrated in Fig.~\ref{fig:feynman_asym}.
In contrast, for $\ell_i H \to \chi^\dagger \phi^\dagger$ and $\ell_i H \to \chi\phi$, there is only an $s$-channel diagram at tree level, with the leading asymmetry arising from the interference between this diagram and loop diagrams with an $s$-channel Majorana neutrino and propagator or vertex loops. These are also shown in Fig.~\ref{fig:feynman_asym}. 
We note that one-loop diagrams with a Majorana neutrino in the $t$-channel do not generate a strong phase and therefore do not contribute to asymmetries at leading order.
More details about calculating the corresponding matrix elements are provided in App.~\ref{appx:Mcalc}. 

The final results of our calculations can be expressed in terms of the expansion coefficients for asymmetric scattering kernels defined in Eq.~\eqref{eq:defexpansionsdelta}.
We find $\lodelta = 0$ for all processes, while for $\nlodelta$ specializing to identical couplings $y_{ia} = y_a$ for all $n_g=3$ SM lepton generations we obtain 
\beq
\nlodelta^{\ell_iH}_{\ell_j^\dagger H^\dagger} &=&
\frac{3}{32\pi^2}\,\mathrm{Im}\!\left(\sum_{a,b,c}\frac{{y_a^*}^2M_a}{M_a^2}\frac{y_b\lambda_bM_b}{M_b^2}\frac{y_c\lambda_c^*}{M_c^2}\right) \ ,
\label{eq:d1a}
\\
\nlodelta^{\ell_iH}_{\chi^\dagger\phi^\dagger} &=&
\frac{1}{32\pi^2}
\mathrm{Im}\!\left(\sum_{b,c,d}\frac{y_b^*\lambda_b^*M_b}{M_b^2}\frac{y_c^*\lambda_c}{M_c^2}\frac{\lambda_d^2M_d}{M_d^2} + \frac{3n_g}{2}\sum_{a,b,c}\frac{y_a^2M_a}{M_a^2}\frac{y_b^*\lambda_b^*M_b}{M_b^2}\frac{y_c^*\lambda_c}{M_c^2}\right) \ ,
\label{eq:d1b}
\\
\nlodelta^{\ell_iH}_{\chi\phi} &=&
\frac{1}{32\pi^2}
\mathrm{Im}\!\left(\sum_{b,c,d}\frac{y_b\lambda_bM_b}{M_b^2}\frac{y_c\lambda_c^*}{M_c^2}\frac{{\lambda_d^*}^2M_d}{M_d^2} + \frac{3n_g}{2}\sum_{a,b,c}\frac{y_a^2M_a}{M_a^2}\frac{y_b^*\lambda_b^*M_b}{M_b^2}\frac{y_c^*\lambda_c}{M_c^2}\right) \ ,
\label{eq:d1c}
\\
\nlodelta^{\chi\phi}_{\chi^\dagger\phi^\dagger} &=&
\frac{n_g}{16\pi^2}\,\mathrm{Im}\!\left(\sum_{b,c,d}\frac{y_b\lambda_bM_b}{M_b^2}\frac{y_c^*\lambda_c}{M_c^2}\frac{{\lambda_d^*}^2M_d}{M_d^2}\right) \ .
\label{eq:d1d}
\eeq
These relations satisfy a number of consistency conditions. From Sec.~\ref{subsec:L} we know that to obtain physical weak phases we need at least two flavors of Majorana neutrinos with different masses. Here, we see that each term within the parentheses in the expressions above is manifestly real when there is only a single heavy neutrino flavor. A second check is that these relations are symmetric under the interchange of $\lambda_a\leftrightarrow y_a$ and swapping visible and hidden sector particles once lepton flavors and SU(2)$_L$ factors are taken into account. 

A more complicated condition on the asymmetric coefficients is implied by CPT and unitarity. Together, these require that for any initial state $\alpha$~\cite{Toussaint:1978br,Weinberg:1979bt,Nanopoulos:1979gx,Kolb:1979qa,Barr:1979wb,Hook:2011tk},
\beq
\sum_\beta\Delta\mathcal{W}^{\alpha}_{\beta}(s) = 0 \ ,
\label{eq:cptu}
\eeq
where the sum runs over all possible final states.\footnote{This relation holds at finite temperature in the MB approximation. There is a related generalization when full quantum statistics is included~\cite{Hook:2011tk,Weinberg:1979bt,Toussaint:1978br}.} 
This condition is fulfilled by the asymmetry coefficients collected in Eqs.~\eqref{eq:d1a}--\eqref{eq:d1d}. Specifically, it is straightforward to check that 
\beq
0 &=& \sum_j\nlodelta^{\ell_iH}_{\ell_j^\dagger H^\dagger} + \nlodelta^{\ell_iH}_{\chi^\dagger\phi^\dagger} + \nlodelta^{\ell_iH}_{\chi\phi} \ ,
\label{eq:cptu1}
\\
0&=& \nlodelta^{\chi\phi}_{\chi^\dagger\phi^\dagger} + \sum_j\nlodelta^{\chi\phi}_{\ell_j^\dagger H^\dagger} 
+ \sum_j\nlodelta^{\chi\phi}_{\ell_j H}  \ ,
\label{eq:cptu2}
\eeq
where in the second relation we have used $\nlodelta^{\chi\phi}_{\ell_j^\dagger H^\dagger} = \nlodelta^{\ell_jH}_{\chi^\dagger\phi^\dagger}$ and $\nlodelta^{\chi\phi}_{\ell_j H} = - \nlodelta_{\chi\phi}^{\ell_j H}$.

\subsection{Overview of Dark Sector Processes}

The second class of new processes in our scenario are reactions entirely within the dark sector through the interactions of Eq.~\eqref{eq:sink}. In contrast to transfer reactions, these are mediated primarily by the light $\eta$ fermion and thus remain active down to low temperatures. Dark sector processes play an important role in determining the ultimate contribution of the dark sector to the cosmological abundance of dark matter.

At high dark temperatures, $T_x \gg m_\phi,\,m_\chi$, the main role of dark sector reactions is to maintain equilibrium among the $\chi$, $\phi$, and $\eta$ states. Indeed, as long as the dimensionless coupling $y_\chi$ of Eq.~\eqref{eq:sink} is not much smaller than unity, energy transferred from the visible sector to the dark sector will quickly be thermalized within the dark sector to a temperature $T_x < T$. We consider $y_\chi \sim 1$ to assume effectively instantaneous dark thermalization throughout our analysis, but will comment further on the size of $y_\chi$ in Sec.~\ref{sec:sink}. 

When $T_x$ falls to near or below $m_{\phi,\chi}$, it is convenient to change from the 2-component fermion formulation we have been using to a 4-component one. We define (in an abuse of notation)
\beq
\chi = \left(\begin{array}{c}\chi\\\chi^{c\,\dagger}\end{array}\right) \ , \qquad
\eta = \left(\begin{array}{c}\eta\\0\end{array}\right) \ .
\eeq
In this notation, the coupling of Eq.~\eqref{eq:sink} translates into
\beq
-\lag_{sink}  = y_\chi\,\phi^*\,\overline{\chi^c}P_L\eta + \mathrm{h.c.}~.
\eeq
The physical dark states consist of a massive complex scalar $\phi$, a lighter Dirac fermion $\chi$, and a massless Weyl fermion $\eta$.
In our study we assume $m_\phi > m_\chi$, although the opposite mass ordering would yield a similar phenomenology. 

As the dark temperature falls below $T_x\sim m_\phi$, the decay $\phi \to \chi  \eta$ dominates over the inverse decay and the density of $\phi$ is depopulated. Crucially to our scenario, these decays preserve dark sector lepton number $L_x$ (as well as $L+L_x$); 
they only transfer any net charge from the $\phi$ population to one in $\chi$. The decay rate is
\beq\label{eq:phi_decay}
\Gamma(\phi \to \chi{\eta}) =
\frac{y_\chi^2}{16\pi}\,m_\phi\!\left[1-\lrf{m_\chi}{m_\phi}^2\right]^2 \ .
\eeq
This is much larger than Hubble when $T_x\sim m_\phi$ for all the parameter values considered in this work.

Evolving further to $T_x \sim m_\chi$, the interactions of Eq.~\eqref{eq:sink} also provide an annihilation channel $\chi \bar{\chi}\to \eta \bar{\eta}$.
This reaction depletes the symmetric density of $\chi$, transferring it to the massless $\eta$ fermions which act as a \emph{dark sink} in the sense of Refs.~\cite{Bhattiprolu:2023akk, Bhattiprolu:2024dmh}. However, since this reaction preserves the charge $L_x$ it does not alter the asymmetry in $\chi$. The scattering kernel for the reaction is
\beq
\mathcal{W}^{\chi\bar{\chi}}_{\eta\bar{\eta}}(s) =
\frac{y_\chi^4}{8\pi}\left[
1 - \frac{2(m_\phi^2-m_\chi^2)}{\sqrt{s(s-4m_\chi^2)}}\,\ln\!\lrf{1+A}{1-A}
+\lrf{m_\phi^2-m_\chi^2}{m_\phi^2-m_\chi^2+s/2}^{\!2}\!\frac{1}{1-A^2}
\right] \ ,
\eeq
where $2A = \sqrt{s(s-4m_\chi^2)}\Big/(m_\phi^2-m_\chi^2+s/2)$.
The corresponding cross section at low velocities $v$ in the center-of-mass frame is
\beq\label{eq:chi_ann_to_eta}
\sigma v &=& \frac{y_\chi^4}{32\pi}\frac{m_\chi^2}{(m_\phi^2+m_\chi^2)^2}\left[1 + \mathcal{O}(v^2)\right] \\
&\simeq&
\left(7.2\times 10^{-21}\,\text{cm}^3/\text{s}\right)
\times \lrf{y_\chi}{0.5}^4\lrf{\text{GeV}}{m_\chi}^2\frac{1}{[(m_\phi/m_\chi)^2+1]^2}\ .
\nnmb
\eeq
This leading term corresponds to the $s$-wave in a multipole expansion and is unsuppressed at the low velocities relevant for $\chi$ freeze-out.
If this cross section is large enough, the relic density of $\chi$ can be dominated by the asymmetry~\cite{Graesser:2011wi, Iminniyaz:2011yp}, corresponding to asymmetric dark matter~\cite{Kaplan:2009ag, Nussinov:1985xr, Petraki:2013wwa, Zurek:2013wia}.

\section{Reheating and Transfer to the Dark Sector}
\label{sec:cosmo}

Having presented an overview of our scenario, we turn next to investigating its cosmological evolution. As described in Sec.~\ref{sec:model}, our starting point is reheating after primordial inflation. We assume that this reheating populates the visible sector nearly exclusively, and produces only negligible abundances of the dark sector and heavy Majorana states.
In this section we compute the rate of energy transfer to the dark sector following reheating of the visible sector. This lays the foundation for our upcoming asymmetry calculations, which we defer to the next section. 
A necessary condition for asymmetry generation in the scenario is $T_x < T$; this imposes constraints on the relative size of $\Trh$ and $M_a$. We also show that for $\Trh \ll M_a$ the direct production of heavy Majorana neutrinos after inflation is small enough to be neglected. 
Finally, we comment on the plausibility of the assumptions we make about primordial reheating.

\subsection{Populating the Dark Sector by Transfer}
\label{subsec:heat_transfer}

Starting from a visible sector at $\Trh$ and an empty dark sector, transfer reactions initiated by visible states will quickly populate the dark sector. These reactions are mediated by heavy neutrinos. If they are fast enough, transfer reactions can even fully equilibrate the two sectors. We compute here the amount of energy transferred to the dark sector and the conditions under which full equilibration is avoided.

For $\Trh \gg m_\phi,\,m_\chi,\,m_Z$, the evolution of the energy densities of the dark and visible sectors can be approximated by 
\beq
\frac{d\rho}{dt} &=& - 4\Hub \rho\phantom{_x} -{{C}}_E \ ,
\label{eq:rhoevol}\\
\frac{d\rho_x}{dt}&=& -4\Hub \rho_x + {{C}}_E \ ,
\nnmb
\eeq
where $\rho$ ($\rho_x$) is the visible (dark) energy density, $\Hub = \sqrt{(\rho+\rho_x)/3\mpl^2}$ is the Hubble parameter with the reduced Planck mass $\mpl \simeq 2.435\times 10^{18} \gev$, and ${{C}}_E$ is the energy transfer collision term. Assuming self-thermalization within both sectors, which is expected given the particle content of each sector, we can relate the energy densities to temperatures by
\beq
\rho = \frac{\pi^2}{30}g_*T^4 \ ,
\qquad
\rho_x = \frac{\pi^2}{30}g_{*x}T_x^4 \ ,
\label{eq:temps}
\eeq
where $g_* \simeq 106.75$ and $g_{*x} \simeq 7.25$ are the effective numbers of relativistic energy degrees of freedom in the visible and dark sectors.

For $T_x\ll T$ and $T \ll M_a$, the energy transfer collision term ${{C}}_E$ is dominated by the ${2\to 2}$ reactions $\ell H\to \chi\phi$, $\ell H\to \chi^{\dagger}\phi^{\dagger}$ and their conjugates.  Energy transfer from other reactions is suppressed by powers of $T_x/T < 1$. The leading contributions from each of these channels can be obtained from Tab.~\ref{tab:Mvalues} together with the results collected in App.~\ref{appx:collterm}. Neglecting also asymmetries between these processes and their conjugates, which are expected to be small, we find 
\beq
{{C}}_E &\simeq&
2[{{C}}_E]^{\ell H}_{\chi^\dagger\phi^\dagger}
+ 2[{{C}}_E]^{\ell H}_{\chi\phi}
-2[{{C}}_E]_{\ell H}^{\chi^\dagger\phi^\dagger}
- 2[{{C}}_E]_{\ell H}^{\chi\phi}
\label{eq:cenergy}
\\
&=& \frac{3}{\pi^4}\losigma^{\ell H}_{\chi^\dagger\phi^\dagger} T^7 + 
\frac{192}{\pi^4}\left(\nlosigma^{\ell H}_{\chi^\dagger\phi^\dagger}+\nlosigma^{\ell H}_{\chi\phi}\right)T^9
-\bigg\{T\to T_x\bigg\} \ ,
\nnmb
\eeq
where the definition for $[{{C}}_E]$ is given in App.~\ref{appx:collterm}. 
This expression, together with the relations of Eq.~\eqref{eq:temps}, provides everything needed to evaluate Eq.~\eqref{eq:rhoevol}.

While our final results are based on a full numerical evaluation of Eq.~\eqref{eq:rhoevol}, it is instructive to obtain an approximate solution valid for $T_x\ll T$. In this limit, the visible sector dominates the total energy density and therefore the Hubble rate. Moreover, energy transfer does not significantly impact the time evolution of $T$, which scales approximately as $T\propto 1/a$. The energy density of the dark sector is then
\beq
\frac{\rho_x}{\rho} \ \simeq \ \bigg[1-\lrf{T}{\Trh}\bigg]\!\lrf{{{C}}_E^{(1)}}{\rho \Hub}_{\!\Trh}
\ + \ \frac{1}{3}\bigg[1-\lrf{T}{\Trh}^{\!3}\bigg]\!\lrf{{{C}}_E^{(2)}}{\rho \Hub}_{\!\Trh} \ ,
\label{eq:rhoxratio}
\eeq
where ${{C}}_E^{(n)}$ is portion of the full collision term obtained from terms $[\mathcal{S}_n]$ in Eq.~\eqref{eq:cenergy}. From this expression we see that energy transfer in this scenario is UV dominated, with the change in $\rho_x$ relative to $\rho$ occurring at temperatures $T$ near reheating. The asymptotic dark sector temperature resulting from this transfer is
\beq\label{eq:xi}
\frac{T_x}{T}\ \equiv \ \xi  \ \simeq \ \left[\frac{g_*}{g_{*,x}}\!\lrf{{{C}}_E^{(1)}+{{C}}_E^{(2)}/3}{\rho \Hub}\right]^{1/4}_{\!\Trh} \ .
\eeq
The self-consistency of this approximate result requires $\xi < 1$. We will see later that $\xi < 1$ is also needed for the generation of asymmetries and to avoid cosmological constraints. In practice, we will encounter values $\xi \sim 0.1$--$0.5$ over the parameter ranges that can produce interesting charge asymmetries, which is consistent with this approximation.

\subsection{Production of Heavy Neutrinos}
\label{subsec:productionheavyN}

Similar methods can be used to compute the density of heavy neutrinos created after reheating. The number density of heavy neutrino species $a$ evolves after reheating as
\beq
\frac{dn_a}{dt} = -3\Hub n_a +{{C}}_n^a \ ,
\eeq
with the number collision term ${{C}}_n^a$ given to a very good approximation by~\cite{Kolb:1979qa}
\beq
{{C}}_n^a &\simeq& \left<\frac{1}{\gamma}\right>_{\!T}\!\Gamma_{N_a\to\ell H}\,\bigg({n}_a^{eq}(T) -n_a\bigg)
\\
&& \ + \ \left<\frac{1}{\gamma}\right>_{\!T_x}\!\Gamma_{N_a\to\chi \phi}\,\bigg({n}_a^{eq}(T_x) -n_a\bigg)
\ ,
\nnmb
\eeq
where
\beq
\Gamma_{N_a\to\ell H} = \sum_i\frac{|y_{ia}|^2}{8\pi}M_a \ , \qquad
\Gamma_{N_a\to\chi \phi} = \frac{|\lambda_{a}|^2}{16\pi}M_a \ ,
\label{eq:gamman}
\eeq
with $\langle1/\gamma\rangle_{T_i} = \langle m_a/E_a\rangle_{T_i}$ is the thermally averaged time dilation factor, ${n}_a^{eq}(T_i)$ is the equilibrium number density of $N_a$ at temperature $T_i$, {and the decay widths implicitly include both the listed decay and its C-conjugate (\emph{e.g.} $\Gamma_{N_a\to \ell H}$ has both $N_a\to \ell H$ and $N_a\to \ell^\dagger H^\dagger$)}.

For the initial conditions $n_a\to 0$, $T_x \to 0$, and $T \to \Trh\ll M_a$ at reheating, the yield $Y_a=n_a/s$ of $N_a$, with $s=2\pi^2g_{*s}T^3/45$ being the visible entropy density, is always less than or equal to the equilibrium value at temperature $M_a \gg T > T_x$,
\beq
Y_a \leq \frac{{n}_a^{eq}(T)}{s} \simeq 
\frac{45}{\pi^2(2\pi)^{3/2}g_{*s}}\!\;x_a^{3/2}e^{-x_a} \ ,
\eeq
where $x_a = M_a/T$. For $x_a \gtrsim 25$, we find that this density of $N_a$ is too small to produce a significant asymmetry of baryons or dark matter. Since we wish to focus on asymmetry creation through scattering, we therefore restrict ourselves to $M_1/\Trh \geq 25$. Moreover, for such large values of $M_1/\Trh$ we also find that the population of dark sector states from thermal $2\to 1$ production of $N_a$ followed by $N_a\to \chi^\dagger\phi^\dagger$ is negligible compared to the direct $2\to 2$ transfer production discussed above.\footnote{This is also needed to justify our truncated expansion of the scattering kernel in powers of $s/M_a^2$.}

While we restrict ourselves to $M_1/\Trh \geq 25$, it should be noted that this model could have a viable parameter space for smaller values of $M_1/\Trh$. In such a regime, the primordial abundance of $N_a$ particles is no longer negligible and our scattering calculations would have to be augmented with the decay of the heavy Majorana neutrinos as well as significant population of the dark sector through decays. As $M_1/\Trh$ decreases to near unity, conventional (asymmetric dark) leptogenesis dominated by decays takes over, as studied in Ref.~\cite{Falkowski:2011xh}, and the scattering contributions become negligible.

\subsection{Comments on Uneven and Instantaneous Reheating}

In this work we assume that inflationary reheating is uneven in that it populates the visible sector to a much higher degree than the dark sector or the heavy neutrinos. This serves two key purposes for the asymmetry creation mechanism we study. First, a small initial dark sector density enables this sector to achieve a different (smaller) temperature than the visible providing a departure from equilibrium. Second, a tiny initial population of heavy neutrinos ensures that the dominant source of asymmetries is scattering rather than the out-of-equilibrium decays of these massive states. Here we argue that our assumptions about reheating can be realized in plausible inflationary scenarios for the ranges $M_1/\Trh \gtrsim 25$ and $\xi = T_x/T \sim 0.1$--$0.5$ that we consider in our asymmetry calculations.

For inflationary reheating to populate the visible sector over the dark sector, the inflaton must decay much more readily to the visible sector. This can be achieved with sufficiently uneven couplings of the inflaton to the two sectors~\cite{Adshead:2016xxj,Hardy:2017wkr,Adshead:2019uwj}.\footnote{Analogously, in preheating scenarios~\cite{Kofman:1994rk,Kofman:1997yn,Felder:1998vq} the visible sector can be populated preferentially if the modes that are excited connect mainly to the SM~\cite{Hardy:2017wkr}.} If the reheating process is nearly instantaneous, this condition is sufficient to ensure a small dark sector density at the reheating temperature $\Trh$. We note that nearly instantaneous reheating can be consistent with current data~\cite{Martin:2024qnn}.

Even if reheating is not instantaneous, our assumptions about the state of the universe after reheating can still be realized to a good approximation. In the standard picture of perturbative reheating, the universe undergoes a phase of domination by inflaton oscillations between the end of inflation and reheating at visible temperature $\Trh$~\cite{Albrecht:1982mp,Dolgov:1982th,Abbott:1982hn}. During this oscillation phase, the visible temperature can be much larger than $\Trh$ which can enhance thermal reactions through higher-dimensional operators~\cite{McDonald:2015ljz,Chen:2017kvz,Kaneta:2019zgw,Garcia:2020eof}, or facilitate the production of states heavier than $\Trh$~\cite{Chung:1998rq,Giudice:2000ex}. In our scenario, temperatures greater than $\Trh$ can enhance the rates of transfer reactions or facilitate direct $N_a$ production before full reheating, and may even lead to a full thermalization of the visible and dark sectors before reheating. However, these effects are countered by the continuous injection of entropy into the visible sector by inflaton decays, which strongly dilutes the reaction products created before the end of reheating. In the end, we find that as long as the two sectors are thermally decoupled at $\Trh$ and that $\xrh = M_1/\Trh$ is large enough, the final population of the dark sector well after reheating is expected to be nearly identical to what would be obtained from instantaneous reheating with dark sector population only after $\Trh$. 

{Calculations supporting the claims above are collected in App.~\ref{appx:reheat}, and we summarize the key results here. 
We model the reheating process with a scalar field oscillating in a quadratic potential and gradually populating the visible SM sector by decays. 
Energy transfer to the dark sector during and after reheating can proceed through $2\to 2$ reactions such as $\ell H \to \chi\phi$ as well as $2\to 1$ production $\ell H\to N_a$ followed by $N_a\to \chi\phi$. 
For $2\to 2$ reactions through higher-dimensional operators, generalizing the results of Refs.~\cite{Chen:2017kvz,Kaneta:2019zgw} for particle number transfer to energy transfer, we find that the transfer is dominated by $T\sim \Trh$ for operator dimensions less than $d<53/6$. 
The contributions from just before the end of reheating are of the same size or less than those from after reheating. 
In our scenario, the dominant transfer comes from $d=5$ operators and thus non-instantaneous reheating gives similar results to instantaneous in this channel. 
For energy transfer through $2\to 1$ production of on-shell $N_a$, there is an enhancement from reactions at temperatures $M_1\gtrsim T \gtrsim \Trh$ relative to after reheating. 
However, the total contribution from such $2\to 1$ processes falls off more quickly with the ratio $\xrh = M_1/\Trh$ than for $2\to 2$ processes with $d=5$ operators.
Therefore the $2\to 2$ contribution dominates for sufficiently large $\xrh$, and for the parameters considered in our analysis we find $\xrh \gtrsim 25$ is typically enough for this. 
Nearly identical arguments can be applied to asymmetry generation.}

{
In addition to thermal reactions during and after reheating, energy could also be transferred to the dark sector or the massive neutrinos by gravitational production~\cite{Ren:2014mta,Garny:2015sjg,Ema:2018ucl,Mambrini:2021zpp}. This source depends on the Hubble rate during inflation, but is subleading relative to the thermal channels considered above for the lighter dark sector states. For the massive neutrinos $N_a$, their gravitational production is suppressed with our working assumption of $M_1/\Trh \gg 1$. Moreover, in contrast to gravitational DM candidates they decay promptly at reheating over the parameter ranges we study and therefore do not accumulate a large density over time.}


\section{Asymmetry Creation in the Early Universe}
\label{sec:BE}

We now turn to study the creation of charge asymmetries in our scenario. The specific charges we track are $B-L$ and $L_x$ since they are broken individually by transfer reactions but are conserved by all other reactions in the theory. 
These charges are generated predominantly at temperatures near $\Trh$, characteristic of UV freeze-in.
In this section, we study in detail the creation of $B-L$ and $L_x$ charges through scattering. First, we connect these charges with individual particle densities by imposing constraints implied by the conservation of other charges as well as equilibrium relations enforced by fast reactions within the visible and dark sectors. Next, we derive the full set of Boltzmann equations for charge creation. After that, we investigate specific solutions to these equations. Finally, we study more broadly the parameter regions over which our scenario can account for the observed baryon abundance, while relating it to the asymmetry generated in the dark sector.

\subsection{Relations Between Particle and Charge Densities}
\label{subsec:chemicals}

Once created, charge asymmetries are quickly redistributed by much faster reactions within each of the visible and dark sectors. We can account for the effect of these fast intra-sector reactions through relations between different  chemical potentials.

Recall that a relativistic particle species $X$ with chemical potential $\mu_X$ has an asymmetric number density of~\cite{Haber:1981fg,Haber:1981ts}
\beq\label{eq:n_DeltaX}
n_{\Delta X} \simeq \frac{1}{6}g_Xk_X\mu_XT_X^2 \ ,
\eeq
where $g_X$ is the number of internal degrees of freedom, $k_X = 1~(2)$ for a fermion~(boson), and $T_X$ is its temperature. Moreover, if a reaction $\alpha \to \beta$ is much faster than Hubble, the chemical potentials of the species involved equilibrate according to~\cite{Kolb:1990vq,Rubakov:2017xzr}
\beq
\sum_{\alpha}\mu_{\alpha} = \sum_\beta\mu_\beta \ ,
\hspace{1.0cm} \Gamma_{\alpha\to\beta} \gg \Hub \ .
\eeq
Finally, given an approximately conserved charge $Q$, the net charge density is related to species number densities by
\beq
n_Q \ \equiv \ \sum_iQ_in_i \ , 
\eeq
where the sum runs over all particle species with charges $Q_i$.

Starting in the visible sector at the high temperatures $T \sim \Trh$ where the asymmetries are predominantly created, we expect unbroken electroweak symmetry and the equilibration of all SM gauge interactions~\cite{Harvey:1990qw}. In this temperature range we also expect the equilibration of both strong and weak sphaleron transitions, as well as most of the Yukawa interactions. In our analysis we assume that all the Yukawa interactions are equilibrated for simplicity.\footnote{This is true for most of the parameter space we study and the error introduced otherwise is small~\cite{Buchmuller:2005eh,Chen:2007fv,Fong:2012buy}.} Imposing the equilibration relations from these reactions on the chemical potentials and demanding zero net hypercharge, we find that all SM chemical potentials can be related to the net $B-L$ charge. Most importantly for us, we obtain
\beq
\mu_{\ell_i}  =  -\frac{7}{79}\,\mu_{B-L} \ ,
\qquad
\mu_H = -\frac{4}{79}\,\mu_{B-L} \ ,
\label{eq:musm1} 
\eeq
where we have defined $\mu_{B-L} \equiv 6n_{B-L}/T^2$. Note that our assumption of equilibrated lepton Yukawa interactions together with equal lepton couplings to the $N_a$ imply that all the lepton asymmetries are equal. 

Below the electroweak symmetry breaking temperature the electroweak sphalerons turn off~\cite{Khlebnikov:1988sr,DOnofrio:2014rug} and the relations of Eq.~\eqref{eq:musm1} are modified. For these lower temperatures, the final baryon density is~\cite{Harvey:1990qw}
\beq
\mu_B = \frac{28}{79}\,\mu_{B-L} \ ,
\eeq
where $\mu_{B} \equiv 6n_{B}/T^2$. 

Turning next to the dark sector, at the high temperatures relevant for asymmetry creation, $T\sim \Trh$, all dark sector states are effectively massless by assumption. We expect that the interaction of Eq.~\eqref{eq:sink} will be equilibrated as well as (in most cases) the Dirac mass of $\chi$. Demanding zero net U(1)$_D$ charge as well, all number densities can be related to the dark sector lepton charge $L_x$ defined in Tab.~\ref{tab:charges}. For $T_x \gg m_\phi,\,m_\chi$ we find
\beq
\mu_\chi  = \frac{1}{2}\,\mu_{L_x} =  -\mu_{\chi^c}
\ , \qquad
\label{eq:muhid1}
\mu_\phi = \frac{1}{2}\,\mu_{L_x}
\ , \qquad
\mu_\eta = \mu_\phi - \mu_\chi =  0 \ ,
\eeq
where $\mu_{L_x} \equiv 6n_{L_x}/T_x^2$. As $T_x$ falls below $m_\phi$, the asymmetry in $\phi$ is transferred to $\chi$ and $\eta$ such that 
\beq
\label{eq:chemLx}
\mu_\chi = \mu_\eta = -\mu_{\chi^c}  = \mu_{L_x} \ . 
\eeq
This relation remains fixed at $T_x < m_\chi$ by U(1)$_D$ conservation. 

The chemical potential relations above allow us to account for the effect of fast intra-sector interactions in our model and relate all final asymmetric number densities to the net $L_x$ and $B-L$ charges produced at high temperatures.

\subsection{Boltzmann Equations for Asymmetries}
\label{subsec:BE_TRH}

Let us now turn to study the generation and time evolution of the charge asymmetries in $B-L$ and $L_x$. These charges are only broken in the theory by transfer reactions. In this section we obtain the Boltzmann equations that govern the creation and redistribution of $B-L$ and $L_x$ asymmetries. 
Some related technical details are collected in App.~\ref{appx:fullBE}. 

The general form of the Boltzmann equations for charges $Q= B-L,\,L_x$ is
\beq
\frac{dn_Q}{dt} + 3\Hub n_Q = {{C}}_Q \ ,
\label{eq:boltzq}
\eeq
where ${{C}}_Q$ is the collision term describing reactions that change the charge. This term follows from the general form of Eq.~\eqref{eq:collision}, and involves a sum of contributions from all relevant reactions $\alpha\to \beta$ weighted by the corresponding charge change $\Delta Q^\alpha_\beta$. Here, these reactions correspond to the transfer reactions in Tab.~\ref{tab:transfer}, together with their charge and time conjugates. In general, ${{C}}_Q$ depends on the visible and dark temperatures $T$ and $T_x$ as well as the charges~$n_Q$. 

Since we expect charge densities to be small relative to total particle number densities during creation, we expand the charge collision terms to linear order in the asymmetries (or equivalently the chemical potentials). Upon summing over all reactions, the charge collision terms can be put in the form
\beq
{{C}}_Q = S_Q - W_Q \ ,
\label{eq:cqsplit}
\eeq
where $S_Q$ is a \textit{source} term that depends on asymmetric matrix elements but is independent of the charge asymmetries themselves, and $W_Q$ is a \textit{wash-out} term that depends on symmetric matrix elements but is also linear in charge asymmetries.

Using CPT and unitarity in the form of Eqs.~(\ref{eq:cptu1},\ref{eq:cptu2}), the source terms for $B-L$ and $L_x$ can be written exclusively in terms of the asymmetries in the reaction rates for $\ell_i H\to \ell_j^\dagger H^\dagger$ and $\chi\phi \to \chi^\dagger\phi^\dagger$, respectively. For $B-L$, we find
\beq
S_{B-L} \ = \ \frac{12n_g^2}{\pi^4}\big[\mathcal{D}_2\big]^{\ell_i H}_{\ell_j^{\dagger} H^{\dagger}}\,(1-\xi^8)\,T^8 
\ \equiv \
\sigma_{B-L}\,\frac{T^8}{M_1^4}
\ ,
\label{eq:sourcebl}
\eeq
where $n_g=3$ is the number of lepton generations, $\xi = T_x/T \leq 1$,
and $\sigma_{B-L}$ is dimensionless. For $L_x$, we get
\beq
S_{L_x} = \frac{12}{\pi^4}\big[\mathcal{D}_2\big]^{\chi\phi}_{\chi^{\dagger} \phi^{\dagger}}\,(1-\xi^8)\,T^8 
\ \equiv \
\sigma_{L_x}\,\frac{T^8}{M_1^4}
\ .
\label{eq:sourcelx}
\eeq
In the latter forms of these expressions, the coefficients $\sigma_Q$ are nearly temperature independent for $\xi \ll 1$. They are also both proportional to asymmetric matrix elements that violate the charges within each sector and require C and CP violation.

A note is in order about these sources in view of the strong constraints on charge creation through scattering derived in Refs.~\cite{Toussaint:1978br,Weinberg:1979bt,Nanopoulos:1979gx,Kolb:1979qa,Barr:1979wb}, and Ref.\cite{Hook:2011tk} in particular, based on equilibration, CPT, and unitarity. Both sources are seen to vanish when $\xi=1$, corresponding to equilibration between the visible and dark sectors. This addresses the primary obstacles in Refs.~\cite{Toussaint:1978br,Weinberg:1979bt,Nanopoulos:1979gx,Kolb:1979qa,Barr:1979wb}. The more recent study of Ref.~\cite{Hook:2011tk} expands these arguments to the case of freeze-in between two sectors at different temperatures to show that CPT and unitarity imply that the transfer of a conserved charge between two sectors is proportional to at least the \textit{third} power of the portal interaction, \textit{i.e.} the \textit{next-to-leading order}, and therefore very suppressed. Our scenario evades this suppression by explicitly breaking the overall $L+L_x$ symmetry. After integrating out the massive $N_a$ neutrinos, the combination $y_a \lambda_a$ can be thought of as the effective freeze-in portal between the two sectors. Taken together, Eqs.~(\ref{eq:d1a},\ref{eq:d1d}) and Eqs.~(\ref{eq:sourcebl},\ref{eq:sourcelx}) demonstrate that our source terms are proportional to the portal coupling to the \textit{second} power, \textit{i.e.} the \textit{leading order}. The key difference relative to Ref.~\cite{Hook:2011tk} comes from insertions of the Majorana masses $M_a$ that break the $L+L_x$ symmetry explicitly. Indeed, by applying CPT and unitarity we have written the sources of Eqs.~(\ref{eq:sourcebl},\ref{eq:sourcelx}) such that they are proportional to the contributions from reactions that break $L+L_x$ explicitly \emph{within} each sector.

The wash-out terms receive contributions from all reactions listed in Tab.~\ref{tab:transfer}, including from processes that do not generate asymmetries. We find 
\beq
\left(\begin{matrix}
W_{B-L}\\W_{L_x}
\end{matrix}\right)
= \left[\frac{T^3}{M_1^2}\!\left(
\begin{matrix}\omega_{B-L,B-L}^{(1)}&\omega_{B-L,L_x}^{(1)}\\\omega_{L_x,B-L}^{(1)}&\omega_{L_x,L_x}^{(1)}\end{matrix}
\right)
+ \frac{T^5}{M_1^4}\!\left(
\begin{matrix}\omega_{B-L,B-L}^{(2)}&\omega_{B-L,L_x}^{(2)}\\\omega_{L_x,B-L}^{(2)}&\omega_{L_x,L_x}^{(2)}\end{matrix}
\right)
\right]
\!\left(\begin{matrix}n_{B-L}\\n_{L_x}\end{matrix}\right) \ ,
\label{eq:wash-out}
\eeq
where 
\beq
\omega_{B-L,B-L}^{(1)} &=& \frac{33 M_1^2}{79\pi^4}n_g\Big[~\Big(
2n_g\losigma^{\ell_iH}_{\ell_j^\dagger H^\dagger}
+2n_g\losigma^{\ell_i\ell_j}_{H^\dagger H^\dagger}
+\losigma^{\ell_iH}_{\chi^\dagger\phi^\dagger}
+\losigma^{\ell_iH}_{\chi\phi}
\Big)
\\
&&~~+ \xi^3\Big(
\losigma^{\ell\chi}_{H^\dagger \phi^\dagger}
+ \losigma^{\ell\chi^\dagger}_{H^\dagger \phi} 
+ \losigma^{\ell\phi}_{H^\dagger \chi^\dagger} 
+ \losigma^{\ell\phi^\dagger}_{H^\dagger \chi} 
\Big)~
\Big] \ ,
\nnmb\\
\nnmb\\ %
\omega_{B-L,L_x}^{(1)} &=& -\frac{3 M_1^2}{\pi^4}n_g\Big[~
\Big(\losigma^{\ell\chi}_{H^\dagger \phi^\dagger}
- \losigma^{\ell\chi^\dagger}_{H^\dagger \phi} 
+ \losigma^{\ell\phi}_{H^\dagger \chi^\dagger} 
- \losigma^{\ell\phi^\dagger}_{H^\dagger \chi} 
\Big)
\\
&&~~+ \xi^3\Big(
\losigma^{\ell H}_{\chi^\dagger\phi^\dagger}
-\losigma^{\ell H}_{\chi\phi}\Big) 
~\Big]  \ ,
\nnmb\\
\nnmb\\ %
\omega_{L_x,B-L}^{(1)} &=& -\frac{33 M_1^2}{79\pi^4}n_g\Big[~
\Big(
\losigma^{\ell_iH}_{\chi^\dagger\phi^\dagger}
-\losigma^{\ell_iH}_{\chi\phi}
\Big)
\\
&&\hspace{0.5cm} + \xi^3\Big(
\losigma^{\ell\chi}_{H^\dagger \phi^\dagger}
- \losigma^{\ell\chi^\dagger}_{H^\dagger \phi} 
+ \losigma^{\ell\phi}_{H^\dagger \chi^\dagger} 
- \losigma^{\ell\phi^\dagger}_{H^\dagger \chi} 
\Big)~
\Big] \ ,
\nnmb\\
\nnmb\\ %
\omega_{L_x,L_x}^{(1)} &=& \frac{3 M_1^2}{\pi^4}\Big[~
n_g\Big(
\losigma^{\ell\chi}_{H^\dagger \phi^\dagger}
+ \losigma^{\ell\chi^\dagger}_{H^\dagger \phi} 
+ \losigma^{\ell\phi}_{H^\dagger \chi^\dagger} 
+ \losigma^{\ell\phi^\dagger}_{H^\dagger \chi} 
\Big)
\\
&&\hspace{0.5cm} + \xi^3\Big(
2\losigma^{\chi\phi}_{\chi^\dagger\phi^\dagger}
+2\losigma^{\chi\chi}_{\phi^\dagger\phi^\dagger}
+n_g\losigma^{\ell_i H}_{\chi^\dagger \phi^\dagger}
+n_g\losigma^{\ell_iH}_{\chi\phi}
\Big) 
~\Big]  \ ,
\nnmb %
\eeq
where $\losigma$ coefficients can be found in Tab.~\ref{tab:Mvalues}. 
Factors of $M_1$ have been added to make the $\omega^{(1)}_{QR}$ terms dimensionless.
The expressions for $\omega_{QR}^{(2)}$ are identical but with the replacements $\{1,\xi^3\}\losigma \to 24M_1^2\{1,\xi^5\}\nlosigma$ for reactions with intra-sector initial states (\emph{e.g.} $\ell_i H$ or $\chi\phi$) or $\{1,\xi^3\}\losigma \to 24M_1^2\{\xi,\xi^4\}\nlosigma$ for inter-sector (\emph{e.g.} $\ell \chi$) initial states.  

A key aspect of these expressions is the overall signs of the various terms. Specifically, they imply that when $n_Q$ is non-zero, the wash-out term for the charge~$Q$ tends to relax it to zero. 
However, we also see that a non-zero charge $n_{Q'\neq Q}$ can act as a \textit{source} for charge~$Q$. Global stability is guaranteed by the feature that wash-out matrices $\omega^{(1,2)}_{QR}$ have non-negative traces and determinants.

The Boltzmann equations for $Q= B-L,\,L_x$ of Eq.~\eqref{eq:boltzq} together with the expressions for the source terms of Eqs.~(\ref{eq:sourcebl},\ref{eq:sourcelx}), the wash-out terms of Eq.~\eqref{eq:wash-out}, and the temperature evolution of Eqs.~(\ref{eq:rhoevol},\ref{eq:temps}) provide a closed system of equations that we solve numerically. To interpret our numerical results, it is instructive to write these equations in the simplifying case when $T_x/T = \xi \ll 1$, where energy transfer from the visible to the dark sector does not significantly alter the time evolution of the visible temperature $T$ and $\Hub \simeq \sqrt{\rho/3\mpl^2}$. This implies approximate conservation of entropy in the visible sector. In turn, we have that 
\beq
u \ \equiv \ \frac{\Trh}{T} 
\eeq
is proportional to the scale factor, $u\propto a(t)$, provided $g_{*s}$ is constant as we expect for $T\gg 100 \gev$. Adopting $u$ as the evolution variable and normalizing the charge densities by the visible entropy density $Y_Q = n_Q/s$ with $s=2\pi^2g_{*s}T^3/45$, we obtain
\beq
\frac{d}{du}\left(\begin{matrix}Y_{B-L}\\Y_{L_x}\end{matrix}\right)
&\simeq& 
\frac{1}{us\Hub}\!
\left(\begin{matrix}S_{B-L}-W_{B-L}\\S_{L_x}-W_{L_x}\end{matrix}\right)
\label{eq:boltzyq}\\
&\simeq&
\frac{1}{u^4}
\,\frac{M_1^4}{s(M_1)\Hub(M_1)}\frac{1}{\xrh^3}\!
\left(\begin{matrix}\sigma_{B-L}\\\sigma_{L_x}\end{matrix}\right) 
\nnmb\\
&&
- \frac{1}{u^2}
\,\frac{M_1}{\Hub(M_1)}\frac{1}{\xrh}\!
\left(
\begin{matrix}\omega_{B-L,B-L}^{(1)}&\omega_{B-L,L_x}^{(1)}\\\omega_{L_x,B-L}^{(1)}&\omega_{L_x,L_x}^{(1)}\end{matrix}
\right)\!
\left(\begin{matrix}Y_{B-L}\\Y_{L_x}\end{matrix}\right)
\nnmb\\
&&
-\frac{1}{u^4}
\,\frac{M_1}{\Hub(M_1)}\frac{1}{\xrh^3}\!
\left(
\begin{matrix}\omega_{B-L,B-L}^{(2)}&\omega_{B-L,L_x}^{(2)}\\\omega_{L_x,B-L}^{(2)}&\omega_{L_x,L_x}^{(2)}\end{matrix}
\right)\!
\left(\begin{matrix}Y_{B-L}\\Y_{L_x}\end{matrix}\right) \ .
\nnmb
\eeq
Here $\xrh \equiv M_1/\Trh$, and $s(M_1)$ and $\Hub(M_1)$ refer to the visible entropy density and Hubble rate in the radiation phase formally evaluated at $T=M_1$ with $\xi=0$.

\subsection{Charge Creation and Dark Wash-In}
\label{subsec:dark_washin}

To study the evolution of charges in our scenario we fix a set of benchmark parameters. It is convenient to do so in terms of ratios relative to $M_1$, $|\lambda_1|$, and $|y_1|$ and a pair of phases. For our primary set of benchmarks, we take
\beq
M_2 = 3M_1 \ , \qquad
y_1 = |y_1|e^{i\alpha} \ , \qquad
y_2 = 2|y_1| \ , \qquad
\lambda_1 = |\lambda_1|\,e^{i\beta} \ , \qquad
\lambda_2 = |\lambda_1| \ .
\label{eq:benchmark}
\eeq 
The phase conventions here are completely general and can be obtained by field redefinitions. We also set the magnitude of the lepton Yukawa coupling $|y_1|$ to be the maximum value allowed by the neutrino mass constraint of Eq.~\eqref{eq:nubound} derived from Eq.~\eqref{eq:seesaw}. {With the benchmarks above, this implies the numerical value
\beq
|y_1|^2 = \frac{2.32\times10^{-3}}{|1+3e^{i\alpha}|}
\bigg(\frac{M_1}{10^{12}\,\gev}\bigg)\bigg(\frac{\Delta m_\nu}{0.07\,\eV}\bigg) \ ,
\eeq
evaluated with $\Delta m_\nu \to 0.07\,\eV$. Note as well that we have ${|y_1|\propto \sqrt{M_1}}$. }
Larger values of the couplings tend to produce greater asymmetries, whereas in the limit $M_2 \gg M_1$ the heavier neutrino decouples and the CP violation needed for the source terms is suppressed.
These benchmark values are therefore optimistic in terms of asymmetry generation, but they are not tuned or reliant on any symmetry.  To fully specify the model, the only remaining quantities to fix are $M_1$, $|\lambda_1|$, and the two phases.

Using the benchmark coupling ratios, we obtain source terms in Eqs.~(\ref{eq:sourcebl},\ref{eq:sourcelx})
\beq
\sigma_{B-L} &=& \frac{12n_g^2}{\pi^4}\,|y_1^4\lambda_1^2|\left[\frac{2}{3}\sin(\alpha-\beta)-\frac{10}{27}\sin(\alpha+\beta)+\frac{32}{27}\sin(2\alpha)\right](1-\xi^8) \ ,
\label{eq:sigbl}
\\
\sigma_{L_x}&=& \frac{12}{\pi^4}\,|y_1^2\lambda_1^4|\left[0 \, \sin(\alpha-\beta)-\frac{16}{27}\sin(\alpha+\beta)+\frac{5}{27}\sin(2\beta)\right](1-\xi^8) \ .
\label{eq:siglx}
\eeq
A crucial point here is that both phases appear in both source terms. This implies that $\sigma_{B-L}/|y_1^4\lambda_1^2|$ and $\sigma_{L_x}/|y_1^2\lambda_1^4|$ are expected to be of similar magnitudes in the absence of tuning. The wash-out coefficients $\omega_{QR}^{(n)}$ can be evaluated in the same way. They have nearly the same form but with cosines of the phases rather than sines, and typically do not vanish as the phases go to zero. 

The dependence of the source and wash-out coefficients on the two phases has a useful implication. When these phases are smaller than unity, rescaling them according to $\alpha\to \zeta\alpha$, $\beta\to \zeta\beta$ with $\zeta < 1$ 
induces $C_E\to C_E$, $\omega_{Q,Q'}^{(n)} \to \omega_{Q,Q'}^{(n)}$, and $\sigma_Q \to \zeta \sigma_Q$ to a very good approximation.  Now, the evolution equations of Eqs.~(\ref{eq:rhoevol},\ref{eq:boltzq},\ref{eq:cqsplit}) are invariant under these transformations if we also shift $\rho\to \rho$, $\rho_x\to \rho_x$, and $Y_Q \to \zeta Y_Q$. It follows that a solution for the densities and asymmetry yields for one set of phases also implies a solution for smaller phases rescaled by $\zeta < 1$ in which the energy densities $\rho$ and $\rho_x$ are unchanged but both charge yields are reduced by a factor of $\zeta$. We will make use of this feature below when we investigate the conditions under which our scenario can generate the observed baryon abundance.

To illustrate the specific impact of the phases on asymmetry generation, we consider four phase benchmark values
of $(\alpha,\beta)$:
\beq
\begin{array}{cclcl}
\text{B1} &=& (-2\pi/3,\pi/4) & \to & \big\{1.21,0.76\big\} \, ,\\
\text{B2} &=& (-\pi/3,\pi/4) & \to & \big\{-1.57,0.39\big\}\, ,\\
\text{B3} &=& (0.5215\pi,\pi/4) &\to & \big\{0.00,0.63\big\}\, ,\\
\text{B4} &=& (-2\pi/3,0.6052\pi) & \to & \big\{1.60,0.00\big\}\, .
\end{array}
\label{eq:phasebench}
\eeq
The quantities in braces on the right are the values of the sums of sines in square parentheses appearing in Eqs.~(\ref{eq:sigbl},\ref{eq:siglx}) for $\sigma_{B-L}$ and $\sigma_{L_x}$, respectively. Benchmark~B1 has equal signs for the source term coefficients while benchmark~B2 has opposite signs. Benchmark~B3 is chosen such that the $B-L$ source coefficient $\sigma_{B-L}$ vanishes while benchmark~B4 has $\sigma_{L_x}=0$. The first two benchmarks are reasonably generic for phases on the order of unity, while the second two benchmarks require specific tuning. 
We solve our system of equations with a range of values of model parameters for these benchmark scenarios.

\begin{figure}[!ttt]
  \begin{center}
    \includegraphics[width = 0.99\textwidth]{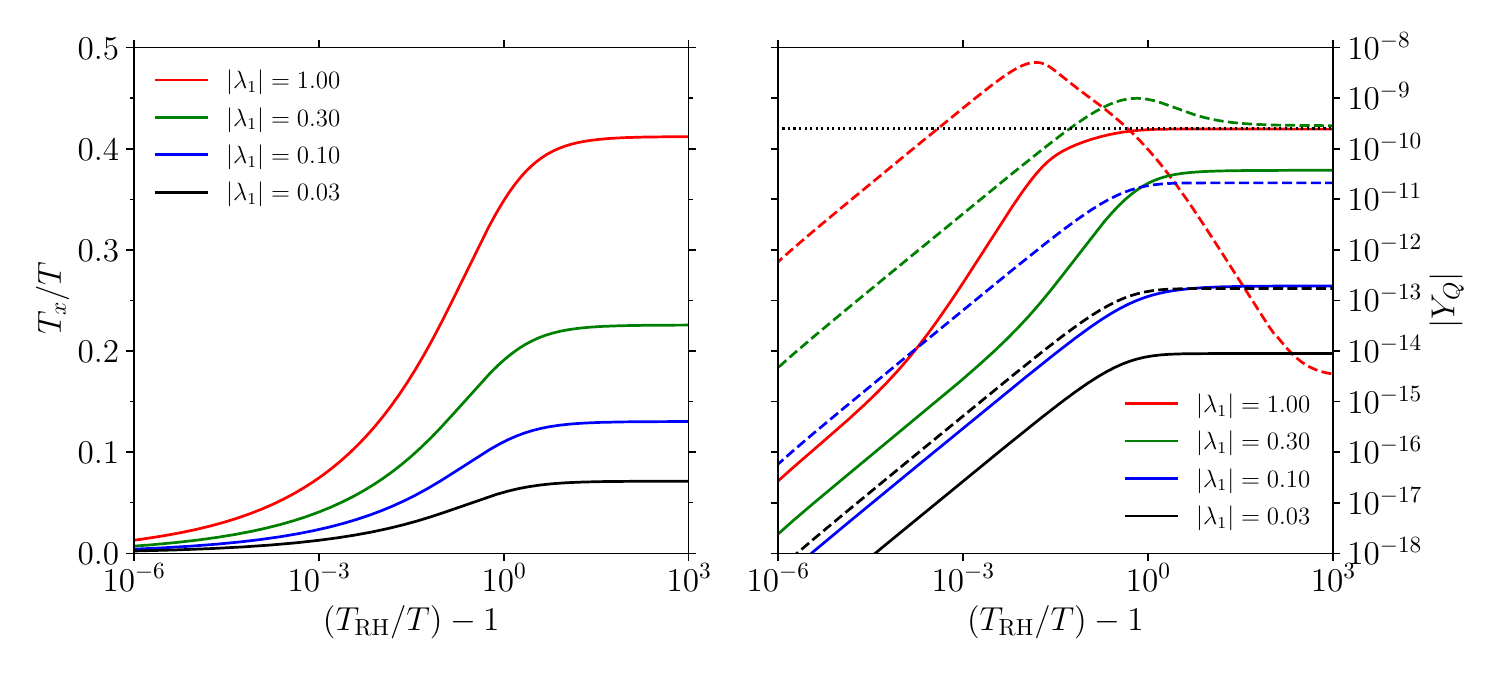}
  \end{center}
  \vspace{-0.5cm}
  \caption{Evolution of the dark temperature ratio $\xi = T_x/T$~(\textbf{left}), and charge densities~(\textbf{right}) of $|Y_{B-L}|$~(solid) and $|Y_{L_x}|$~(dashed) as functions of the inverse temperature relative to reheating for several values of $|\lambda_1|$. Other parameters are specified in the text, with $M_1=10^{10} \gev$, phase benchmark~B1, and $|y_1|$ as large as possible subject to the upper bound on neutrino masses. The dotted black line corresponds to the $|Y_{B-L}|$ value required for the observed baryon asymmetry today. 
  The evolution in the \textbf{left} plot is generic to UV freeze-in models and shows that for the chosen parameters the two sectors do not thermalize~($\xi < 1$). 
  The \textbf{right} plot highlights the possibility of \textit{dark wash-in} 
  at larger $\lambda_1$ values, whereby some of the asymmetry in the dark sector is transferred to the visible sector via wash-out reactions.
  }
  \label{fig:yevol1}
\end{figure}

In the left panel of Fig.~\ref{fig:yevol1} we show the evolution of the dark temperature ratio $\xi = T_x/T$ as a function of the visible temperature parameter $\Trh/T-1$ for the phase benchmark~B1 with $M_1 = 10^{10} \gev$ {(giving $\abs*{y_1} \simeq 2 \times 10^{-3}$)} and several values of $|\lambda_1|$. We also fix the reheating temperature to $\xrh = M_1/\Trh = 30$. As expected from our estimates of Eqs.~(\ref{eq:rhoxratio},\ref{eq:xi}), we find that the temperature ratio $\xi$ grows from zero to a nearly constant value once $\Trh/T - 1 \gtrsim 1$, with the asymptotic value scaling as  
\begin{equation}
\xi \sim \left( \frac{|y_1\lambda_1|^2\mpl}{\xrh M_1}\right)^{1/4} \ .
    \label{eq:xiasymptotic}
\end{equation}
We find as well that when $\xrh$ is held fixed and $|y_1|$ is chosen to saturate the neutrino mass constraint of Eq.~\eqref{eq:nubound}, the asymptotic temperature ratio is nearly independent of $M_1$. For all cases considered in Fig.~\ref{fig:yevol1}, $\xi$ remains well below unity and therefore the dark and visible sectors do not fully equilibrate.

In the right panel of Fig.~\ref{fig:yevol1} we show the evolution of the magnitudes of the $B-L$~(solid) and $L_x$~(dashed) charge yields as a function of $\Trh/T$ for the same benchmark parameters. 
The shapes of different curves can be understood from the charge evolution equations in the approximate form of Eq.~\eqref{eq:boltzyq}. 
Immediately after reheating, the source terms dominate and both charges begin to grow. If the charges become large enough, wash-out reactions can slow or even reverse their growth. The wash-out of one charge can also act as a source for the other through the off-diagonal wash-out coefficients $\omega_{Q,Q'\neq Q}^{(1,2)}$, a process termed \textit{wash-in} in Refs.~\cite{Domcke:2020quw,Domcke:2022kfs,Mojahed:2025vgf}. Most of this evolution occurs within a few Hubble times after reheating due to the overall $1/u^4$ factors in the source terms and the $1/u^2$ factors in the leading wash-out coefficients, which imply that these driving terms eventually decouple. Correspondingly, the charge yields in Fig.~\ref{fig:yevol1} are seen to approach constant values for $u = \Trh/T-1 \gg 1$.

For the parameter ranges considered, we typically have $|\lambda_1| > |y_1|$. This leads to the dark $L_x$ source term, which scales as $\sigma_{L_x}\propto |y_1^2\lambda_1^4|$, being larger than the visible $B-L$ source that scales as $\sigma_{B-L}\propto |y_1^4\lambda_1^2|$. Thus, the dark charge $L_x$ grows more quickly than $B-L$ at early times. However, these larger couplings also enable greater wash-out. Most of the dark wash-out is controlled by the coefficient $\omega_{L_x,L_x}\sim |y_1^2\lambda_1^2|+\xi^3|\lambda_1|^4$. We find that dark wash-out is \emph{strong}, corresponding to a significant decrease in the final charge relative to the source alone, when the following condition holds:
\beq
\left.\frac{M_1}{\Hub(M_1)}\frac{\omega^{(1)}_{L_x,L_x}}{\xrh}\right|_{(u-1)=1} \gg 1 \ .
\label{eq:washxx}
\eeq
This condition is met in our scenario for larger $|\lambda_1|$ and smaller $M_1$. The impact of strong wash-out can be seen in the right panel of Fig.~\ref{fig:yevol1}, specifically in the $|Y_{L_x}|$ curves for $|\lambda_1|=1$ and $|\lambda_1|=0.3$ where at later times we see a distinctive fall off in $|Y_{L_x}|$. Correspondingly, there is an increase in slope for the $|Y_{B-L}|$ curves from wash-in contributions to the charge $B-L$ from the dark sector, in addition to the direct $B-L$ source.
We call this effect \textit{dark wash-in} to highlight the role of the dark sector asymmetry as the primary source of the SM asymmetry in certain parameter regions. 
In contrast, we do not find a strong wash-out of $B-L$ charge due to the constraint on $|y_1|$ from neutrino masses.

\begin{figure}[ttt]
  \begin{center}
    \includegraphics[width = 0.99\textwidth]{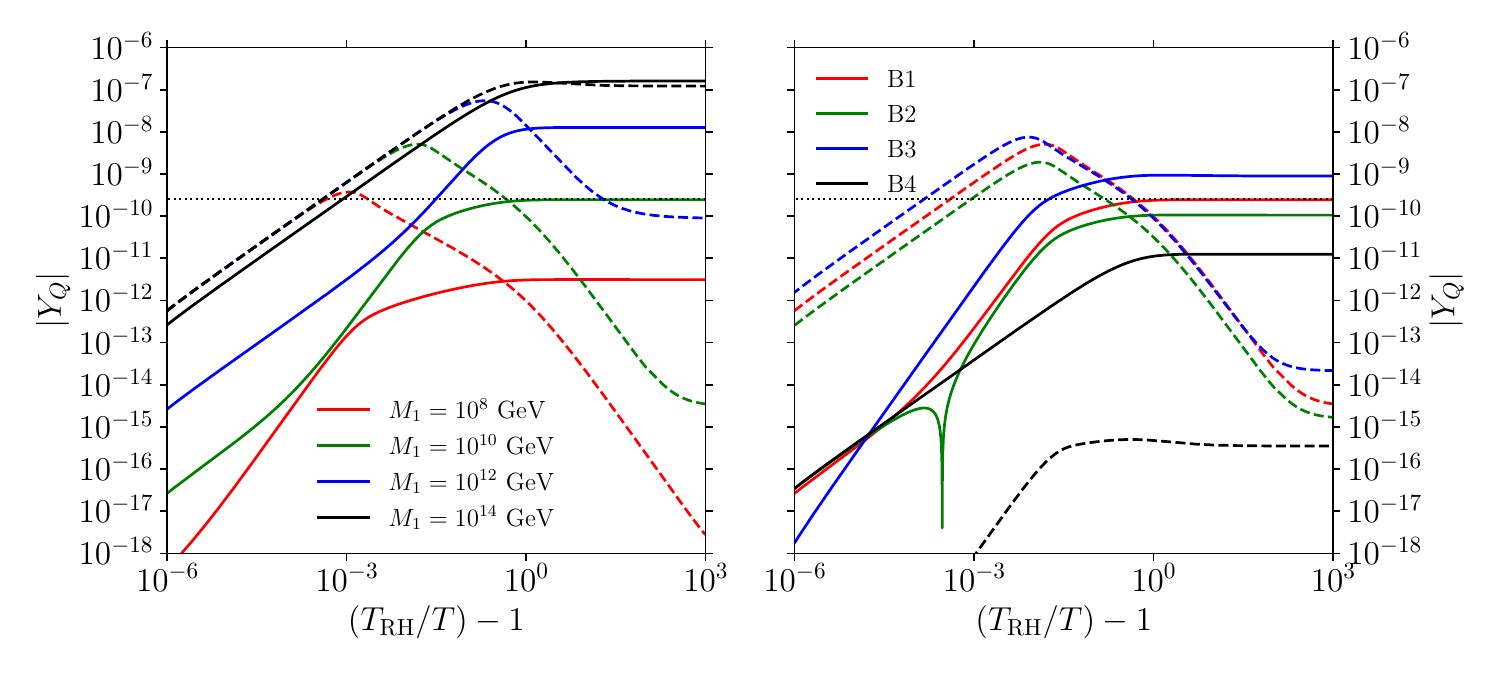}
  \end{center}
  \vspace{-0.5cm}
  \caption{Evolution of the charge densities $|Y_{B-L}|$~(solid) and $|Y_{L_x}|$~(dashed) as a function of the inverse temperature relative to reheating. In the \textbf{left} panel we show curves for $M_1=10^{8,10,12,14} \gev$ and phases as in benchmark~B1, see Eq.~\eqref{eq:phasebench}. In the \textbf{right} panel we show curves for phase benchmarks B1, B2, B3, B4 with $M_1=10^{10} \gev$. For both panels, we also set $\xrh=30$, $|\lambda_1|=1$, and $|y_1|$ as large as possible subject to the neutrino mass bound. 
  The horizontal dotted black line corresponds to the $|Y_{B-L}|$ value required for the observed baryon asymmetry today. 
  The \textbf{left} panel shows that increasing $M_1$ reduces
  wash-out in the dark sector and increases the asymmetries in both sectors.
  The \textbf{right} panel shows that generic benchmarks~B1 and B2, as well as benchmark B3 with zero visible sector source terms, 
  give rise to comparable asymmetries via the dark wash-in mechanism. 
  In benchmark B4 with no dark source term, there is no dark wash-in, but instead a transfer of asymmetry from the visible to the dark sector. 
  }
  \label{fig:yevol2}
\end{figure}

In the left panel of Fig.~\ref{fig:yevol2} we show the effect of varying $M_1$ on charge generation with fixed $|\lambda_1|=1$ for phase benchmark~B1 with $\xrh=30$ and maximal $|y_1|$. 
If $|y_1|$ and $\xrh$ are held fixed, the source and wash-out terms can be shown to both scale as $\mpl/M_1$ at leading order. However, here we adjust $|y_1|$ to be as large as allowed by the neutrino mass constraint of Eq.~\eqref{eq:nubound} which fixes $|y_1| \propto \sqrt{M_1}$ {(with values ranging between $\abs*{y_1} \simeq 2 \times 10^{-4}$ and $0.2$)}. The net effect is that lowering $M_1$ (with fixed $\xrh$ and maximal $|y_1|$) leaves the $L_x$ source unchanged, enhances $L_x$ wash-out, and weakens the $B-L$ source. These effects are evident in the left panel of Fig.~\ref{fig:yevol2}. All the $|Y_{L_x}|$ curves in the plot follow the same trajectory until significant wash-out begins, with the onset of wash-out starting earlier for smaller $M_1$. For $|Y_{B-L}|$, the decrease in the source for lower $M_1$ is also manifest. Another feature of the $|Y_{B-L}|$ curves is that the contribution from dark wash-in becomes increasingly important for smaller direct $B-L$ sources.

The right panel of Fig.~\ref{fig:yevol2} shows the evolution of the charge yields for the four phase benchmarks B1--B4. For B1, the contributions to $Y_{B-L}$ from the source and dark wash-in add constructively. In contrast, for B2 the source and dark wash-in interfere destructively, and produce a dip in $|Y_{B-L}|$ when the contributions from the direct source are overtaken by those from wash-in. In benchmark~B3, the $B-L$ source is tuned to zero (see Eq.~\eqref{eq:phasebench}), and the entire $B-L$ charge yield comes from dark wash-in.\footnote{The final value of $|Y_{B-L}|$ for B3 is higher than for B1 and B2 since the specific phases of B3 lead to a moderate cancellation in the neutrino mass contribution of Eq.~\eqref{eq:seesaw}, allowing a larger value of $|y_1|$.} In benchmark~B4, the phases are tuned to produce a vanishing source for $L_x$, and thus the flow of wash-in is in the opposite direction of the other benchmark scenarios. The entire $B-L$ charge now comes from the source alone, while a much smaller $L_x$ charge is produced by wash-in from the visible sector.

While we have focused on specific ratios of $M_2/M_1$, $|y_2/y_1|$, and $|\lambda_2/\lambda_1|$ (as specified in Eq.~\eqref{eq:benchmark}) in the discussion above, we have checked that moderate variations in these values do not change our qualitative results and primary conclusions. 
In contrast, the more extreme limits of $M_2/M_1 \to \infty$ or $\lambda_2\to 0$ lead to very different behavior. 
The former corresponds to decoupling the $N_2$ neutrino, in which case we are left with effectively only one heavy neutrino that is not able to sustain the CP violation needed for generating asymmetries. In the latter limit, the dark sector states couple exclusively to $N_1$ so that no physical phase remains in the dark sector and the dark source term vanishes (while the SM source term is still non-zero and can generate a $B-L$ asymmetry). As a result, this limit is similar to phase benchmark~B4 from Eq.~\eqref{eq:phasebench}.

\subsection{Asymmetry Yields}
\label{subsec:ayield}

Having gained an understanding of how the visible and dark charges are created, we now compute the values of the visible and dark charges today that can be obtained in our scenario and investigate whether it can explain the observed baryon abundance. Similar to the previous section, we focus on representative benchmark scenarios that capture the generic behavior of relevant quantities across our parameter space. Even though we compute the yields of $B-L$ and $L_x$ charges at temperatures well above the electroweak scale and the masses of the $\chi$ and $\phi$ dark states, their values do not change once the transfer reactions discussed above decouple. Moreover, it is straightforward to relate these charge yields to the final baryon and dark matter~($\chi$) asymmetries using the relations obtained in Sec.~\ref{subsec:chemicals}: $Y_B = (28/79)Y_{B-L}$ and $Y_{\Delta\chi} = Y_{L_x}$.

\begin{figure}[ttt]
  \begin{center}
    \includegraphics[width = 1.00\textwidth]{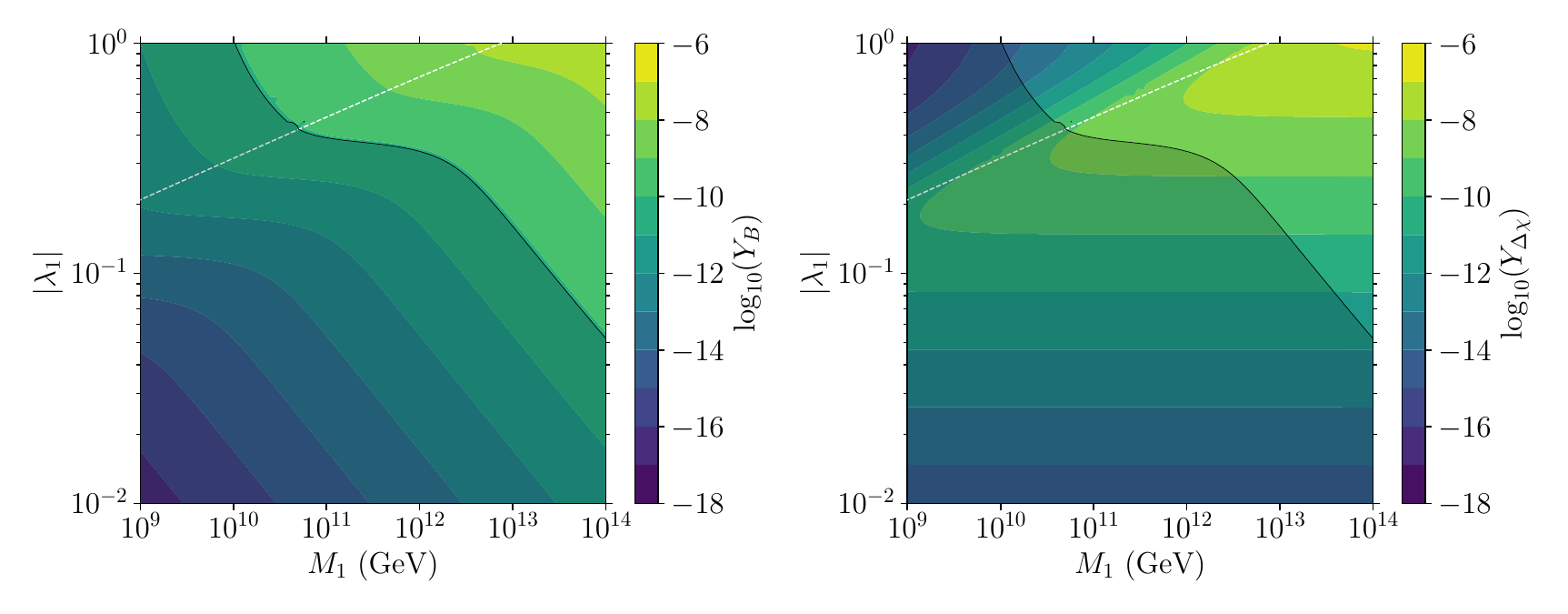}
  \end{center}
  \vspace{-0.5cm}
  \caption{Late time charge densities $Y_B$~(\textbf{left}) and $Y_{\Delta\chi}$~(\textbf{right}) as functions of $M_1$ and $|\lambda_1|$ obtained in the scenario with maximal lepton coupling $|y_1|$, phase benchmark~B1, and $\xrh=30$. The shaded regions in the lower left of both plots fail to produce the observed baryon density and are excluded. The dashed white line  marks the boundary above which the strong dark wash-out condition of Eq.~\eqref{eq:washxx} is met. 
  The \textbf{left} panel shows that in our setup we can produce the full baryon asymmetry for a wide range of parameters. 
  The \textbf{right} panel demonstrates that both asymmetries tend to be similar in magnitude except in the strong dark wash-out regime where the dark charge can be orders of magnitude lower.
  }
  \label{fig:scanml}
\end{figure}

In Fig.~\ref{fig:scanml} we show the late time yields of baryon number $Y_B$ and $Y_{\Delta\chi}$ in the $M_1\text{--}|\lambda_1|$ plane for maximal $|y_1|$, phase benchmark~B1, and $\xrh = 30$. The shaded regions in the lower left of both plots fail to produce the observed baryon asymmetry, $Y_B\simeq 8.7 \times 10^{-11}$, and are excluded~\cite{Planck:2018vyg,ACT:2025fju}. In the upper right unshaded region, the baryon yield for phase benchmark~B1 is greater than the observed value. However, the correct baryon abundance can be obtained at any point in this region with smaller values of the phases $\alpha$ and $\beta$ relative to the benchmark, as discussed in Sec.~\ref{subsec:dark_washin} above.  

The dynamics of charge creation studied in the previous subsection can be seen in Fig.~\ref{fig:scanml}. For $Y_B$, production comes primarily from the direct source term in the region at larger $M_1$ and smaller $|\lambda_1|$. Moving to smaller $M_1$ and larger $|\lambda_1|$, additional $B-L$ is generated by wash-in from the dark sector, corresponding to the contour feature seen in this region. Similarly, for $Y_{\Delta\chi} = Y_{L_x}$ the $L_x$ source term dominates dark charge production for larger $M_1$ and smaller $|\lambda_1|$. As noted in the previous subsection, the source is nearly independent of $M_1$ for maximal $|y_1|$ in this regime. In contrast, in the upper left region of the plot the final dark charge density falls off rapidly with smaller $M_1$ and larger $|\lambda_1|$, and coincides with the onset of strong wash-out of $L_x$. To demonstrate this point, the dashed white line in the plots denotes the boundary above which the strong wash-out condition of Eq.~\eqref{eq:washxx} is satisfied. 

Comparing the two panels in Fig.~\ref{fig:scanml}, we also see that the baryon and dark matter asymmetries can be similar when wash-out is weak. 
On the other hand, in the strong wash-out regime, the dark asymmetry is severely depleted and some of it is transferred to the visible sector.
It was shown previously~\cite{Falkowski:2011xh} that similarly large hierarchies are possible in dark leptogenesis models that rely on heavy neutrino decays where the two sectors are in equilibrium. Our results extend this conclusion to models where the main source of asymmetry-generation is scattering processes and the visible and the dark sectors never reach an equilibrium. In the next section, we argue that this large hierarchy translates into a wide range of viable DM masses that can explain the observed DM abundance today.

\begin{figure}[ttt]
  \begin{center}
    \includegraphics[width = 0.99\textwidth]{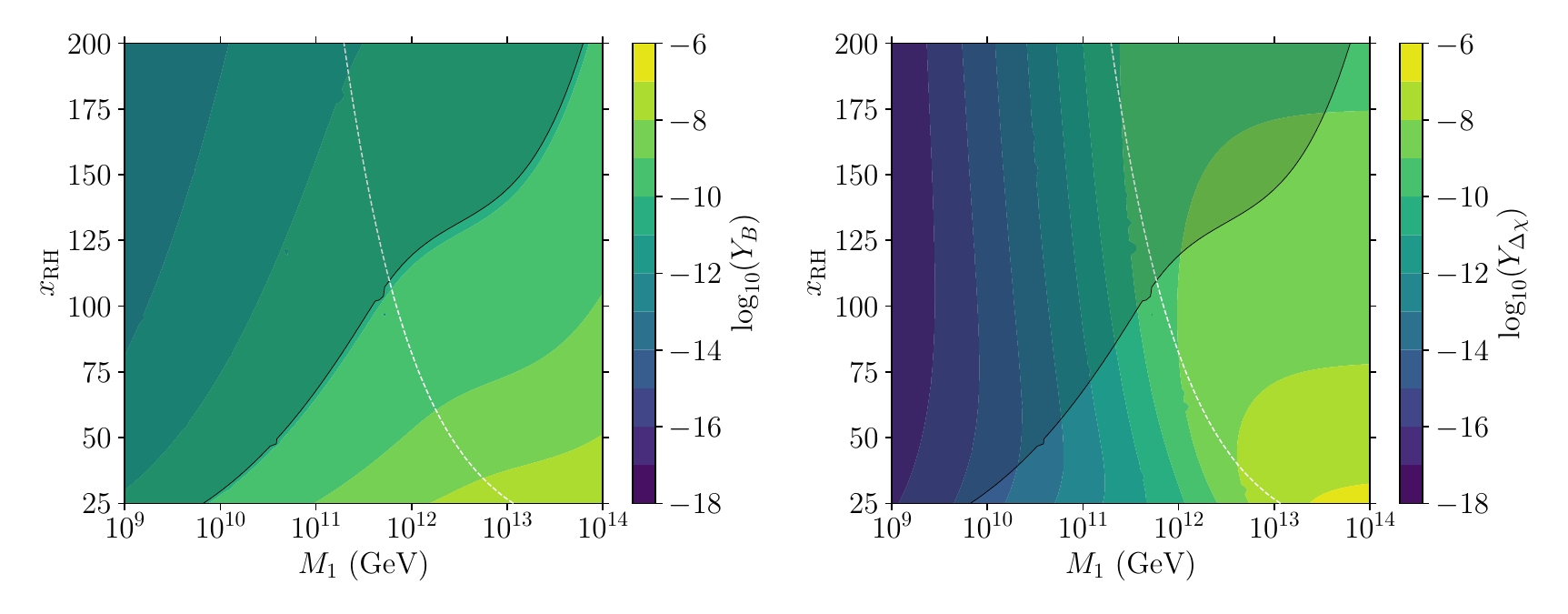}
  \end{center}
  \vspace{-0.5cm}
  \caption{Late time charge densities $Y_B$~(left) and $Y_{\Delta\chi}$~(right) as functions of $M_1$ and ${\xrh = M_1/\Trh}$ obtained in the scenario with maximal lepton coupling $|y_1|$, $|\lambda_1|=1$, and phase benchmark~B1. The shaded regions in the upper left of both plots fail to produce the observed baryon density and are excluded. The strong dark wash-out condition of Eq.~\eqref{eq:washxx} is met to the left of the dashed white line in both panels.
  The \textbf{left} panel shows that by going to smaller $\xrh$ values the observed asymmetry can be attained for lower reheating temperatures. The \textbf{right} panel re-emphasizes the fact that when the wash-out terms are very strong (to the left of the dashed white line) they efficiently transfer the asymmetry generated in the dark sector to the visible sector and give rise to a large hierarchy between the two asymmetries.
  }
  \label{fig:scanmt}
\end{figure}

In Fig.~\ref{fig:scanmt} we show contours of the late time charge densities $Y_B$ and $Y_{\Delta\chi}$ for a range of values of $M_1$ and reheating ratio $\xrh = M_1/\Trh$. In both plots we fix maximal $|y_1|$, $|\lambda_1|=1$, and again use phase benchmark~B1. The shaded regions in both panels show where the predicted baryon density is too low and are excluded. To the lower right of the shaded region the baryon asymmetry is too large, but as discussed above the observed asymmetry can be obtained here with smaller phase values. To the left of the dotted white line in both panels, the strong dark wash-out condition of Eq.~\eqref{eq:washxx} is satisfied and correspondingly we see a reduced dark matter asymmetry.

From the left panel of Fig.~\ref{fig:scanmt} we see that $Y_B$ decreases for larger $\xrh$ when other quantities are held fixed. This is consistent with our expressions for charge evolution in Eq.~\eqref{eq:boltzyq}, where we find that the source terms fall off as $1/\xrh^3$ and the leading wash-out terms decrease as $1/\xrh$. 
Indeed, at larger $M_1$, where the direct $B-L$ source dominates, we see a faster decrease relative to regions at lower $M_1$, in which the dominant source of baryons is wash-in from the dark sector. Related features appear in the right panel for $Y_{\Delta\chi}$. Here, a rapid fall with $\xrh$ is seen at larger $M_1$, where wash-out is weak and the dark source dominates. In contrast, at smaller $M_1$, where the strong dark wash-out regime is realized, the reduction in wash-out for large $\xrh$ counteracts the decrease in the source leading to dark charge densities that are nearly independent of $\xrh$. Note that similar to Fig.~\ref{fig:yevol1}, we find that the DM asymmetry can be orders of magnitude below the SM asymmetry in the strong wash-out regime.

Our results in Figs.~\ref{fig:scanml}-\ref{fig:scanmt} also indicate a lower bound on the reheating temperature $\Trh \gtrsim 4\times 10^8 \gev$ 
if our scenario is to explain the full baryon asymmetry with the benchmark parameters considered. It is interesting to compare this requirement to the Davidson-Ibarra bound derived in Ref.~\cite{Davidson:2002qv}, where the authors showed that thermal leptogenesis models with a type-I seesaw mechanism can only generate the observed baryon asymmetry for reheating temperatures ${\Trh \gtrsim 10^9 \gev}$. Both bounds have the same underlying origin. Smaller $\Trh$ require lower values of $M_1$ to activate the asymmetry generation dynamics. In turn, this implies that smaller lepton Yukawa couplings are needed for consistency with neutrino mass requirements. Since the $B-L$ sources in both scenarios scale as $|y|^4$, the net result is that sufficient asymmetry generation requires a reasonably large reheating temperature. Despite these similarities, we also note that the dark wash-in effect allows us to reach slightly lower reheating temperatures.\footnote{Previous wash-in models have found even lower reheating temperatures~\cite{Domcke:2020quw, Mojahed:2025vgf}, but they do not rely on heavy neutrinos for $B-L$ or CP violation.} 

While the benchmark parameters fixed in Eqs.~(\ref{eq:benchmark},\ref{eq:phasebench}) and studied above are fairly representative of the model parameter space, it is interesting to consider whether deviating from these values could allow us to generate the observed SM asymmetry with lower reheat temperatures. To answer this question, we have performed a broader scan over model parameters to search for the lowest possible reheating temperature $\Trh$ that still generates the observed baryon asymmetry, while remaining consistent with constraints such as the neutrino mass bound of Eq.~\eqref{eq:nubound} and $\xrh > 25$. 
With one significant exception, we find that the lowest possible reheating temperatures for a given $M_1$ value lie within an order-one factor of the values shown in Fig.~\ref{fig:scanml} for benchmark~B1. The specific exception to this conclusion occurs when the couplings and phases are tuned to minimize the contribution to the active neutrino masses via Eq.~\eqref{eq:seesaw}. With such a tuning, much larger values of $|y_1|$ are allowed leading to greatly increased $B-L$ sources, and reheating temperatures as low as $\Trh \sim 10^6 - 10^7\gev$ are possible.  However, given the tuning required to achieve these lower reheat temperatures, we do not study this possibility any further.


\section{Dark Sector Phenomenology}
\label{sec:sink}

After freeze-in and asymmetry creation shortly after reheating, the visible and dark sectors evolve independently. As the universe cools to dark temperatures $T_x \lesssim m_\phi$, the decay ${\phi \to \chi\eta}$ transfers the entire $L_x$ asymmetry to $\chi$. Cooling further to $T_x \lesssim m_\chi$, the annihilation ${\chi \bar{\chi}\to \eta \bar{\eta}}$ depletes the symmetric abundance of $\chi$ but does not alter the asymmetry. At very late times, the dark sector consists of a relic density of dark matter consisting of $\chi$ (and $\bar{\chi}$) together with a dark radiation component from the massless $\eta$ fermions. 

In this section we study the cosmological implications of the dark sector and its asymmetry. We investigate the contribution of $\eta$ to the total radiation density as characterized by its contribution to the effective number of neutrinos, $\Delta\Neff$. We also compute the depletion of $\chi$ through annihilation and determine the range of masses and couplings that allow it to make up the entire dark matter abundance given the asymmetries computed in the previous section. This is found to occur with primarily asymmetric $\chi$ dark matter for masses between $0.1 \gev \lesssim m_\chi \lesssim 10^3 \gev$, while an even greater range of masses is possible for mainly symmetric dark matter.

\subsection{Light Degrees of Freedom}
\label{sec:neff}

To ensure consistency with cosmological observations, particularly the Cosmic Microwave Background~(CMB), the contribution of the $\eta$ fermion to the radiation density near recombination must not be too large. The key parameter here is the change in the effective number of neutrinos, $\Delta\Neff$, which quantifies the total contribution of new relativistic degrees of freedom to the energy density in units of a single neutrino species in the instantaneous decoupling limit. Within the standard neutrino decoupling scenario, $\Neff^\text{SM} \simeq 3.046$, accounting for non-instantaneous neutrino decoupling~\cite{Mangano:2005cc,Cielo:2023bqp}. 
The latest Planck measurement finds $\Neff = 2.99 \pm 0.17$, leading to a $2\sigma$ upper limit of $\Delta\Neff \lesssim 0.3$~\cite{Planck:2018vyg}. Upcoming cosmological surveys such as CMB-S4 aim to further tighten this constraint to $\Delta\Neff \lesssim 0.06$~\cite{CMB-S4:2016ple}. 

The contribution of the Weyl fermion $\eta$ to $\Delta\Neff$ at recombination is
\beq\label{eq:xiCMB}
\Delta\Neff = 
\xi_{\CMB}^4{\lrf{11}{4}^{4/3}} \ ,
\eeq
where $\xi_{\CMB}$ is the ratio of the dark temperature to the visible photon temperature at the epoch of CMB formation.
For the current~(future) constraint $\Delta N_\text{eff} \lesssim 0.3$~($\Delta N_\text{eff} \lesssim 0.06$), this implies $\xi_\CMB \lesssim 0.53$~($\xi_\CMB \lesssim 0.35$).

\begin{figure}[ttt]
  \begin{center}
    \includegraphics[width = 0.65\textwidth]{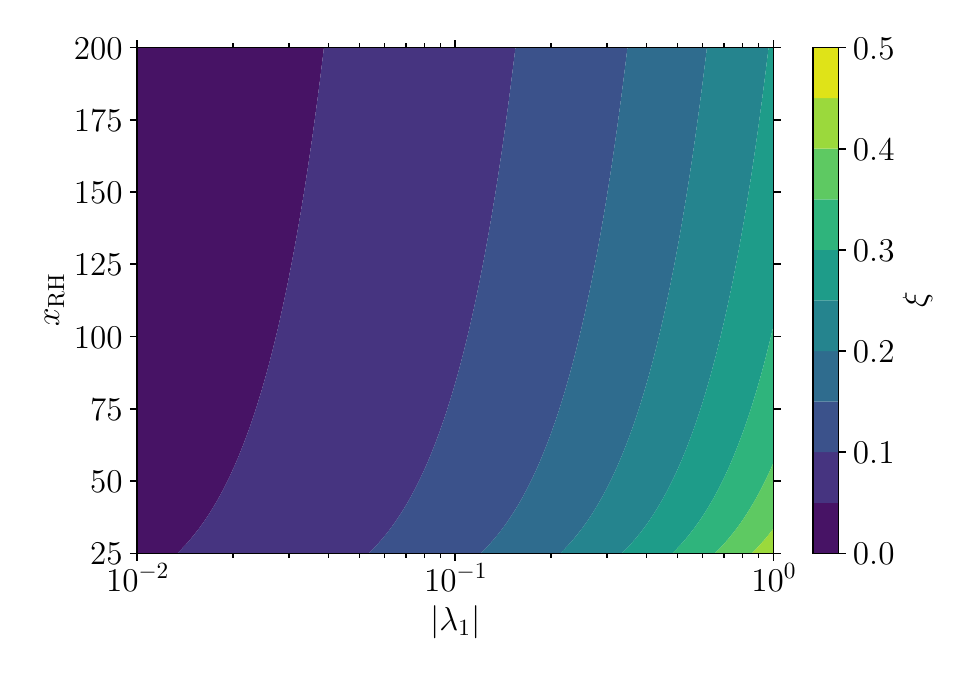}
  \end{center}
  \vspace{-0.5cm}
  \caption{Ratio of the temperatures in the two sectors $\xi = T_x/T$ immediately after the dark sector is populated near reheating for benchmark~B1 in terms of the coupling $|\lambda_1|$ and $\xrh = M_1/\Trh$. We set $M_1= 10^{10}\gev$ in this figure. Since $\xi < 1$ throughout the figure, the dark and visible sectors do not equilibrate with each other anywhere in the parameter regions we study. We find that  everywhere in this parameter space $\xi$ is small enough to evade the bounds from the contribution of $\eta$ to $\Delta N_{\mathrm{eff}}$, see the text for details.}
  \label{fig:ltxi}
\end{figure}

We can relate $\xi_{\text{CMB}}$ to the previously-computed temperature ratio $\xi$ generated by freeze-in at high temperatures ($T,\,T_x \gg m_\phi,\,m_\chi,\,m_Z$) using the separate conservation of entropy in both sectors once transfer reactions have decoupled. This gives
\beq
\xi(T) = 
\xi(T_{\rm FI})\left[
\frac{g_{*s,x}(T_{\rm FI})}{g_{*s}(T_{\rm FI})}\frac{g_{*s}(T)}{g_{*s,x}(T)}\right]^{1/3} \ ,
\label{eq:xifunc}
\eeq
where $T_{\rm FI}$ is the visible temperature after freeze-in has completed and $T_x(T)\equiv \xi(T)\,T$ is treated as a function of $T$. Since we consider freeze-in occurring at very high temperatures, we have the boundary values $g_{*s}(T_{\rm FI}) \simeq 106.75$ and $g_{*s,x}(T_{\rm FI})\simeq 7.25$. Specializing to the temperature of recombination, where we expect the only relativistic degrees of freedom in the dark sector to be the $\eta$ Weyl fermion, we also have $g_{*s,x}(T_{\rm CMB})\simeq 1.75$. This translates into a constraint on the freeze-in temperature ratio $\xi = \xi(T_{\rm FI})$ of
\begin{align}\label{eq:xi_DeltaNeff}
    \xi 
    \lesssim 
    \begin{cases}
        0.99~; & \Delta N_\text{eff}<0.3 \\
        0.66~; & \Delta N_\text{eff}<0.06
    \end{cases}
    \ .
\end{align}
This upper bound on $\xi$ can be compared to the temperature ratios we obtain in our asymmetry calculations. In Fig.~\ref{fig:ltxi} we show $\xi$ as a function of $|\lambda_1|$ and $\xrh = M_1/\Trh$ for benchmark~B1 with $M_1 = 10^{10}\gev$ (but the result is independent of $M_1$). 
We find that over the entire parameter space, the dark sector temperature is low enough to evade current and anticipated future limits on $\Delta N_{\text{eff}}$. 

{
Before moving on, let us comment on the interactions of our theory given in Eq.~\eqref{eq:L}, and specifically the Higgs portal operator $\kappa|\phi|^2|H|^2$ that we did not include among them. Adding this coupling can have an important effect on the effective number of light degrees of freedom in the scenario since it connects the dark and visible sectors without a suppression by the heavy neutrino masses and can lead to their thermalization. Being renormalizable, the thermalization reactions mediated by this operator grow increasingly important relative to the Hubble rate at lower temperatures until the onset of electroweak symmetry breaking, at which point they become highly suppressed. Depending on the size of the Higgs portal coupling $\kappa$, there are three regimes: i) $\kappa$ is very small and the operator does not lead to any significant thermalization of the two sectors, ii) $\kappa$ is moderately small such that the Higgs portal is inactive during asymmetry generation near $T\sim \Trh$ but thermalizes the two sectors with each other at a later time, iii) $\kappa$ is large enough to keep the dark and visible sectors in thermal contact all the way up to reheating. Only regimes i) and ii) are consistent with our asymmetry generation mechanism, and so far we have assumed to be in regime i).}

{
Now, in our full theory with heavy neutrinos, the Higgs portal coupling is generated by log-divergent loops involving the other couplings.\footnote{{Note that no such contribution is generated if the theory is embedded into supersymmetry.}}  Below the heavy neutrino scale, the Higgs portal operator is only renormalized proportionally to itself. This implies that the Higgs portal coupling $\kappa$ can in principle be tuned to near zero at the heavy scale and will stay that way at lower scales, realizing regime i). However, without tuning the natural magnitude of the coupling is $|\kappa| \gtrsim |y\lambda|^2/(4\pi)^2$. For the parameter values considered in our analysis, we find that this loop-generated natural value of $\kappa$ is typically too small to thermalize the two sectors at reheating but large enough to equilibrate them at lower temperatures and continuing down to the weak scale. This corresponds to regime ii) described above, and is consistent with the generation of asymmetries but modifies our estimates above for $\Delta N_{\rm eff}$. To compute the effect, we note that thermalization down to electroweak symmetry breaking in the early universe corresponds to effectively replacing the high-scale freeze-in value of $\xi$ with unity. Using Eq.~\eqref{eq:xifunc}, this leads to a prediction of $\Delta N_{\rm eff} \simeq 0.3$, right on the boundary of what is allowed by current data. Therefore extending our theory to include a natural value of $\kappa$ leads to a prediction for $\Delta N_{\rm eff}$ that is expected to be tested by future surveys~\cite{CMB-S4:2016ple}.
}

\subsection{Dark Matter in a Dark Sink}
\label{subsec:sink}

After freeze-in of the dark sector and decay of the $\phi$ state, the DM candidate $\chi$ has an asymmetric yield $Y_{\Delta \chi}$ as well as a much larger symmetric yield $Y_{\Sigma \chi}$. 
If the symmetric yield is not depleted, it is nearly always too large to allow $\chi$ to be the dark matter. However, our scenario contains the massless $\eta$ fermion into which the $\chi$ state can annihilate. 
These particles form a relativistic thermal bath that absorbs the energy density of the symmetric $\chi$ population, acting as a \emph{dark sink} for this energy~\cite{Bhattiprolu:2023akk, Bhattiprolu:2024dmh}. The dark bath is analogous to the SM photon bath, and does not contribute to the DM density due to the masslessness of its $\eta$ constituents.

The relic abundance of $\chi$ is determined by the symmetric yield $Y_{\Sigma\chi}$ following the freeze-out of the annihilation reaction $\chi \bar{\chi} \to \eta \bar{\eta}$. Computing this relic yield proceeds much like standard dark matter freeze-out~\cite{Gondolo:1990dk,Edsjo:1997bg} but with two twists: the dark temperature $T_x$ differs from the visible one~\cite{Feng:2008mu,Chu:2011be}, and the $\chi$ population has an asymmetry~\cite{Graesser:2011wi,Iminniyaz:2011yp}.\footnote{We assume here without loss of generality that $\chi$ has a positive asymmetry, $Y_{\Delta\chi}\geq 0$.} 
The evolution equations for the $\chi$ and $\bar{\chi}$ densities are
\beq
\frac{dn_\chi}{dt} + 3\Hub n_\chi = 
\frac{dn_{\bar{\chi}}}{dt} + 3\Hub n_{\bar{\chi}}
= - \langle\sigma v\rangle\Big(n_\chi n_{\bar{\chi}} - n_\chi^{eq}n_{\bar{\chi}}^{eq}\Big) \ ,
\label{eq:freezeout}
\eeq
where $n_{\chi,\bar{\chi}}^{eq}=g_\chi(m_\chi T_x/2\pi)^{3/2}\exp[-(m_\chi\mp\mu_\chi)/T_x]$ and $\langle\sigma v\rangle$ are evaluated at the dark temperature $T_x$. 
It is convenient to change variables to
\beq
x_x = \frac{m_\chi}{T_x} \ , \qquad
Z_{\chi,\bar{\chi}} = \frac{x_x^3}{m_\chi^3}n_{{\chi},\bar{\chi}} \ , 
\eeq
and define $Z_{\Delta} = Z_\chi-Z_{\bar{\chi}}$. For $x_x \gtrsim 3$ the only remaining relativistic dark species is $\eta$ so that $g_{*s,x}\simeq 1.75$ is effectively constant. Together with Eq.~\eqref{eq:freezeout}, this implies that $Z_\Delta$ is constant.
The evolution of $Z_{\bar{\chi}}$ in terms of these variables is
\beq
\frac{dZ_{\bar{\chi}}}{dx_x} = -\frac{m_\chi^3\langle\sigma v\rangle}{x_x^4\Hub}\Big(Z_{\bar{\chi}}^2+Z_{\bar{\chi}}Z_\Delta - Z_{eq}^2\Big) \ 
\label{eq:admfo}
\eeq
with $Z_{eq} = g_\chi x_x^{3/2}e^{-x_x}/(2\pi)^{3/2}$. 

To evaluate this equation, we also need the visible temperature $T$ in terms of $x_x$ to determine the Hubble rate, and a relation between $Z_{\chi,\bar{\chi}}$ and $Y_{\chi,\bar{\chi}}$ to set the initial condition. 
The first follows from Eq.~\eqref{eq:xifunc}, but now with $T$ to be taken as a function of $T_x$.
For the second, we have
\beq
Z_{\chi,\bar{\chi}}(x_x) = \frac{2\pi^2}{45}\,g_{*s}(x_x)\,\xi^{-3}(x_x)\,Y_{\chi,\bar{\chi}}(x_x) \ .
\eeq
Note that both of these relations are implicit and can be solved iteratively.

We solve Eq.~\eqref{eq:admfo} using standard techniques~\cite{Graesser:2011wi,Iminniyaz:2011yp} with the freeze-in values of $\xi$ and $Y_{\Delta\chi}$ computed previously as inputs. The dark sector parameters $y_\chi$, $m_\chi$, and $m_\phi$ do not impact freeze-in provided the coupling $y_\chi$ is large enough to self-thermalize the dark sector and the masses $m_\phi,\,m_\chi$ are much smaller than $T_x$ during freeze-in. For our calculations, we set $y_\chi =1$ to ensure dark thermalization while remaining in the perturbative regime, and $m_\phi \geq 1.5\,m_\chi$ to avoid coannihilation effects. Together with the masses, these parameters fix the annihilation cross section given in Eq.~\eqref{eq:chi_ann_to_eta}, which has a leading $s$-wave term at low collision velocities. 

Given an asymmetry $Y_{\Delta \chi}$, the total yield of $\chi$ and $\bar{\chi}$ is bounded below by the asymmetry. In turn, this implies an upper bound on the $\chi$ mass of ${m_\chi \leq m_n(\Omega_{\text{DM}}/\Omega_B)(Y_B/Y_{\Delta \chi})}$, where $m_n$ is the effective nucleon mass. The upper bound is approached when dark matter is almost completely asymmetric, corresponding to strong $\chi \bar{\chi}\to \eta \bar{\eta}$ annihilation. When the annihilation is less efficient, such as for smaller values of $y_\chi$ or larger ratios $m_\phi/m_\chi$, the relic abundance is less asymmetric and the dark matter mass must be smaller to achieve the observed relic density. 

\begin{figure}[ttt]
  \begin{center}
  \includegraphics[width = 0.8\textwidth]{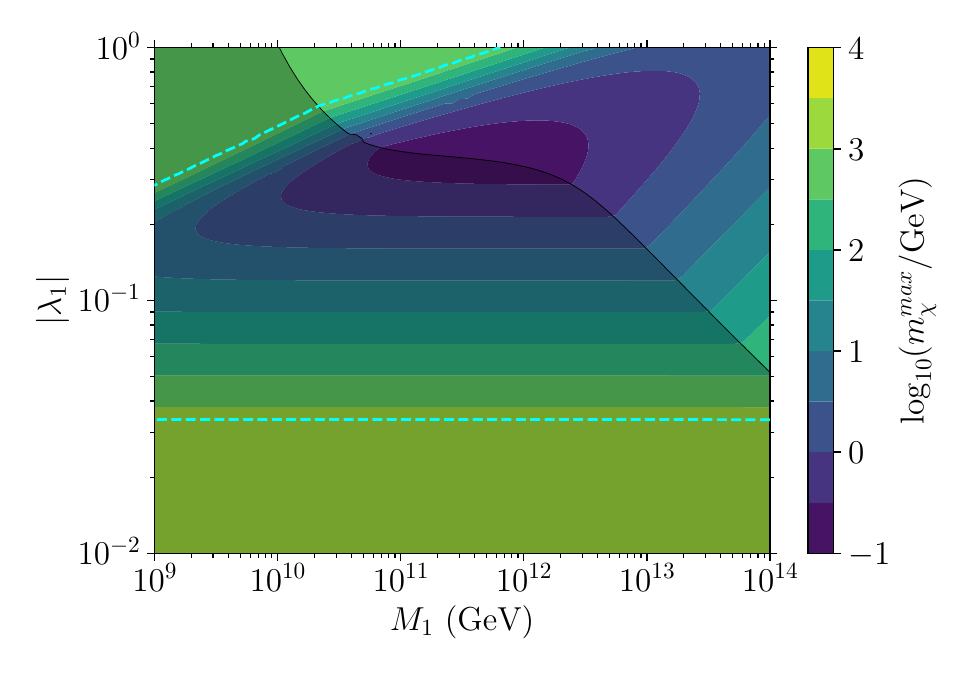}
  \end{center}
  \vspace{-0.5cm}
  \caption{Maximum allowed dark matter mass to obtain the observed dark matter relic density subject to the constraint of generating the full baryon abundance in the $M_1$--$|\lambda_1|$ plane for (rescaled) phase benchmark~B1 with dark matter annihilation to the dark sink with $y_\chi=1$ and $m_\phi=1.5\,m_\chi$. In the shaded region the baryon abundance is smaller than the observed value. In the unshaded region the baryon asymmetry with benchmark~B1 is too large; here we implicitly reduce the weak phases to get the correct baryon abundance while maintaining the ratio $Y_B/Y_{\Delta \chi}$. 
  Between the cyan dashed lines, the dark matter relic population is mainly asymmetric 
  with $Y_{\Delta\chi} \geq 0.8 Y_{\Sigma\chi}$. We find that our scenario can account for the observed dark matter abundance with dark matter masses spanning a wide range.
  }
  \label{fig:madm}
\end{figure}

In Fig.~\ref{fig:madm} we show the maximum allowed $\chi$ dark matter mass to obtain the correct relic density in the $M_1$--$|\lambda_1|$ plane with asymmetries computed using the benchmark~B1 (shown in Fig.~\ref{fig:scanml}) up to phase rescalings and $y_\chi = 1$, $m_\phi=1.5\,m_\chi$. In the shaded region, the baryon asymmetry $Y_B$ is smaller than the observed value $Y_B^{obs}$ and the dark matter asymmetry we use to compute relic densities is precisely that displayed in the right panel of Fig.~\ref{fig:scanml}. 
In the unshaded region, the baryon asymmetry for phase benchmark~B1 is larger than the measured value. 
Here, we apply the weak phase rescaling procedure discussed in Sec.~\ref{subsec:dark_washin} in which we implicitly reduce the phases $\alpha$ and $\beta$ relative to the benchmark to obtain $Y_B = Y_B^{obs}$. 
This procedure also reduces the dark matter asymmetry $Y_{\Delta\chi}$ compared to what is shown in Fig.~\ref{fig:scanml}, but maintains the ratio $Y_B/Y_{\Delta\chi}$. 
The dashed cyan lines in the figure indicate the boundary of the region where the relic abundance can be mostly asymmetric with $Y_{\bar{\chi}}/Y_\chi < 0.1$ (corresponding to $Y_{\Delta\chi} \simeq 0.8 Y_{\Sigma\chi}$). 

Dark matter in this scenario can be mainly asymmetric throughout most of the region where the observed baryon density is obtained. Depending on the relative sizes of the baryon and dark matter asymmetries, maximum (asymmetric) dark matter masses in the range $0.1 \gev \lesssim m_{\chi}^{max} \lesssim 10^3 \gev$ can be achieved. The exception to this occurs in the upper left unshaded region. Referring to Fig.~\ref{fig:scanml}, this region coincides with strong wash-out of the dark charge asymmetry. On account of the wash-out, the dark matter asymmetry $Y_{\Delta\chi}$ is very small here, and larger $\chi$ masses $m_\chi \gtrsim 10^3 \gev$ would be needed to obtain the full dark matter abundance. However, for such large masses the annihilation cross section of Eq.~\eqref{eq:chi_ann_to_eta}, which scales as $1/m_\chi^2$, becomes too small to reduce the symmetric component of the $\chi$ density down to the level of the asymmetry. In this region the dark matter abundance must be primarily symmetric and the maximum dark matter mass plateaus to symmetric freeze-out values for our choice of parameters $y_\chi=1$ and $m_\phi = 1.5\,m_\chi$. 

By going to larger $m_\phi/m_\chi$ values, the annihilation cross section can be significantly reduced allowing for a larger symmetric dark matter component.\footnote{Reduced annihilation cross sections can also be done with smaller $y_\chi$, but when this coupling becomes too small our assumption of rapid dark sector self-thermalization at freeze-in may break down.} 
This allows us to explain the dark matter abundance today with even lower $\chi$ masses. We verify that by varying these parameters we can explain the observed dark matter abundance today down to $m_\chi \sim \keV$ at which point various astrophysical constraints become relevant~\cite{Sigurdson:2009uz,Das:2010ts}.
{Let us also mention that if late-time thermalization between the dark and visible sectors occurs due to the Higgs portal operator $\kappa|H|^2|\phi|^2$ as discussed in Sec.~\ref{sec:neff}, we would have a larger value of $\xi$ during the freezeout process compared to the analysis above. This would tend to increase the final dark matter abundance, and therefore lead to somewhat smaller maximal dark matter masses than shown in Fig.~\ref{fig:madm}.}

Unfortunately, the prospects for testing the dark matter in the model experimentally look bleak. Our dark matter particle $\chi$ is stable and only connects to visible matter through very massive singlet neutrinos with $M_1 \gtrsim 10^{10} \gev$, naively
evading direct detection or collider probes. 
And since the $\chi$ dark matter annihilates to the invisible $\eta$ particles, no useful indirect detection signal is expected either. One might also expect dark matter self-interactions or dissipation involving the massless $\eta$ particles. However, the possible low-energy operators are constrained by the symmetries of the theory. The leading $\chi$--$\eta$ and $\chi$--$\chi$ scattering interactions emerge from four-fermion operators suppressed by $1/m_\phi^2$, and are point-like and too feeble to have an observable impact~\cite{Tulin:2017ara}. 

As a final remark, let us connect our dark sink freeze-out calculations to the stringent unitarity constraints on freeze-in cogenesis models~\cite{Toussaint:1978br,Weinberg:1979bt,Nanopoulos:1979gx,Barr:1979wb,Hook:2011tk}, and comment on how our scenario evades them. 
CPT and unitarity imply cancellations in generating a baryon asymmetry from equilibrium scattering when summed over all production channels~\cite{Nanopoulos:1979gx}. 
These arguments were further extended to the case of two sectors that are out of equilibrium with each other (while maintaining intra-sector equilibrium)~\cite{Hook:2011tk}. Specifically, Ref.~\cite{Hook:2011tk} showed that the transfer of any conserved charge shared between the two sectors is vanishing at leading order in the portal coupling and can only occur at the third power of this coupling. For freeze-in mechanisms, which require a feeble portal coupling, this implies a grossly suppressed asymmetric abundance compared to the symmetric number density. 
In dark matter-baryon cogenesis models, if the dark matter asymmetry is within a few orders of magnitude of the baryon asymmetry, this implies the symmetric dark matter abundance will be much too large unless it is significantly depleted.

Our scenario avoids these pitfalls in two ways. First, the cancellation proven in Ref.~\cite{Hook:2011tk} is avoided in our setup because the shared conserved charge $L+L_x$ is explicitly broken by the heavy neutrino masses $M_a$, and we are able to obtain comparatively larger asymmetries. Indeed, after applying CPT and unitarity we find that our leading source terms are proportional to asymmetries in reactions that explicitly break the shared $L+L_x$ charge, see Eqs.~(\ref{eq:sourcebl},\ref{eq:sourcelx}). 
Second, the dark sink in our model enables an efficient depletion of the symmetric abundance of DM into $\eta$ fermions, which in turn allows us to explain the observed DM abundance with primarily asymmetric dark matter.

\subsection{Other Potential Observables}

{
The primary asymmetry creation dynamics in our scenario must occur at a very high energy scale relative to laboratory experiments, in analogy to minimal leptogenesis, and this makes the scenario difficult to test experimentally. Even so, theoretical considerations such as neutrino mass generation and quantum naturalness motivate potential additional signatures in our mechanism. We comment here on two classes of possibilities.
}

{
Just like standard leptogenesis, the interactions in our scenario can be extended to a theory of neutrino masses and mixings~\cite{Buchmuller:2005eh,Chen:2007fv,Fong:2012buy}. While the mapping from the CP violating phases in the full theory responsible for asymmetry generation to observable phases at low energies is not unique, there is a general expectation for CP violation to arise in the Pontecorvo-Maki-Nakagawa-Sakata~(PMNS) matrix~\cite{Maki:1962mu,Pontecorvo:1967fh} that could potentially be observed at Hyper-Kamiokande~\cite{Hyper-Kamiokande:2018ofw} and DUNE~\cite{DUNE:2020ypp}. Furthermore, our asymmetry generation mechanism favors larger absolute neutrino masses~\cite{Hambye:2003rt}, which can be probed in cosmology~\cite{Lesgourgues:2006nd,DESI:2024mwx} and searches for neutrinoless double beta decay~\cite{Schechter:1981bd,Dolinski:2019nrj,Agostini:2022zub}.
}

{
Our baryogenesis mechanism also relies on a new scalar field $\phi$ that connects with the massive neutrinos. These neutrinos must be very heavy and the scalar-neutrino coupling needs to be large for asymmetry generation, while the scalar must be considerably lighter to enable the efficient depletion of the symmetric dark matter component through annihilation, Eq.~\eqref{eq:chi_ann_to_eta}. Together, these requirements imply large quantum corrections to the dark scalar mass that must be tuned away to keep it relatively light, analogous to corrections to the Higgs mass from heavy seesaw neutrinos~\cite{Vissani:1997ys,deGouvea:2014xba}. This conclusion can be avoided if our scenario is part of a supersymmetric theory with a scale of soft supersymmetry breaking in the dark sector not too far above the weak scale. Interestingly, such an embedding would also suppress a Higgs portal coupling between the dark scalar and the SM Higgs that could lead to the thermalization of the two sectors, as discussed in Sec.~\ref{sec:neff}. On the one hand, this would eliminate the $\Delta N_{\rm eff}$ observable discussed above. On the other, a supersymmetrization of the theory would also lead to potentially observable superpartners in the visible sector.}

\section{Conclusions}
\label{sec:conclusion}

In this work, we have developed a simple model 
for the creation of baryon and dark matter asymmetries through UV freeze-in scattering.
Given the remarkable similarity between the DM and SM matter abundances, 
theories that can explain this concordance
dynamically are more natural candidates for the particle nature of DM and well-motivated targets for various experimental programs. 
Thus, as one of the first viable models of UV freeze-in DM, our results provide more credence for this DM mechanism.

Our setup is an asymmetric DM model through a neutrino portal, with an additional new ultra-light fermion in the dark sector. 
Unlike most models of heavy neutrino leptogenesis, in our UV freeze-in model the universe 
{is assumed to reheat to a temperature well below the masses of the heavy neutrinos with no significant heavy neutrino population, and}
thus the asymmetry can not be generated through the portal neutrino decays. Nonetheless, we show that the 2-to-2 scattering processes mediated by the heavy neutrinos are sufficient for generating enough asymmetry. The large Majorana mass of the neutrino portal particles breaks an extended lepton number. This enables different net lepton numbers between the two sectors, as a result of which DM mass can span a wide range. 

Only a small subset of transfer processes between the two sectors contribute to populating the dark sector as source terms, see Tab.~\ref{tab:transfer}. While the remaining processes do not directly source an asymmetry in dark matter abundance, they have an indispensable role as wash-out terms that can produce qualitatively new effects. After populating the dark sector through the source terms, the wash-out terms can redistribute the generated dark asymmetry to a lepton asymmetry in the visible sector. This is a manifestation of the general idea of wash-in, in which wash-out terms reprocess a primordial asymmetry in some conserved charge into a ${B-L}$ asymmetry. In our model this primordial asymmetry is in the form of a dark matter asymmetry, so we dub our mechanism \textit{dark wash-in}. 

Although this redistribution allows us to circumvent the Davidson-Ibarra bound on thermal leptogenesis models~\cite{Davidson:2002qv}, other cosmological considerations prevent the reheat temperature to be substantially below the original Davidson-Ibarra bound. 
Depending on model parameters (couplings and heavy neutrino masses), we can explain the observed SM and DM~abundances today for $\Trh \gtrsim 10^8 \gev$ and $ 0.1 \gev \lesssim m_\chi \lesssim 10^3 \gev$ with an asymmetric dark matter abundance, while even lower masses are viable for symmetric dark matter abundance.

{In conjunction}
with the asymmetric DM abundance, we also produce a symmetric DM abundance that is orders of magnitude higher and has to be depleted to evade stringent astrophysical bounds on 
the DM density.
To that end, we introduced new ultra-light dark fermions that act as a dark thermal bath into which the DM's symmetric abundance can annihilate. This is a manifestation of the dark sink idea and can be used to avoid overclosing the universe and other astrophysical bounds. It also allows us to explain the visible and dark abundances today with dark matter masses as low as $\mathcal{O}(1) \keV$ for symmetric dark matter abundance.

Our model is a stepping stone towards a complete solution of the coincidence problem that relies on the UV freeze-in mechanism. 
However, to fully solve the coincidence problem, our generation of asymmetric number densities should be infused with a dynamical selection of the DM mass such that the total energy densities of the two sectors are naturally within $\mathcal{O}(1)$ factors of each other. 
Our setup can nevertheless serve as an extreme case, delineating the space of models that could explain the coincidence problem through freeze-in scattering. 
While detecting direct signals of our model appears to be challenging, it would be interesting to explore potential impacts on neutrino properties~\cite{Agostini:2022zub} or inflationary observables~\cite{Cui:2021iie}.

\section*{Acknowledgments}

We thank Joe Bramante, Hooman Davoudiasl, Keisuke Harigaya, Graham Kribs, David McKeen, Alex Millar, Keith Olive, Brian Shuve, Tracy Slatyer, Flip Tanedo, and James Unwin for illuminating discussions. 
The work of P.A. is supported by the U.S. Department of Energy under grant number DE-SC0011640. 
M.M. is supported by subMIT at MIT Physics, the Fonds de Recherche du Québec – Nature et Technologies (FRQNT) doctoral research scholarship (Grant No.~305494), and the Natural Sciences and Engineering Research Council (NSERC) Canada Postgraduate Scholarship - Doctoral (Grant No.~577851), the U.S. Department of Energy, Office of Science, Office of High Energy Physics of U.S. Department of Energy under grant Contract Number DE-SC0012567.
This work is supported by Discovery Grants from NSERC. TRIUMF receives federal funding via a contribution agreement with the National Research Council (NRC) of Canada. M.S. is supported by the US Department of Energy under award number DE-SC0008541.
This research was supported in part by grant NSF PHY-2309135 to the Kavli Institute for Theoretical Physics (KITP).

\appendix

\section{Collision Term Calculations}
\label{appx:collterm}

In this appendix we collect some technical details on calculating the contributions to collision terms appearing in the Boltzmann equations for number and energy densities from $2\to 2$ transfer processes. We follow in the footsteps of Refs.~\cite{Gondolo:1990dk,Edsjo:1997bg}, but generalize their results to reactions involving particles belonging to separate thermal baths with different temperatures. We also follow the calculation of Ref.~\cite{Elahi:2014fsa} to show that in our setup, and given the functional form of our matrix elements, the collision terms have simple specific scalings with temperature.
In these calculations, we will also work within the Maxwell-Boltzmann~(MB) approximation and focus on the specific case of effectively massless particles.

\subsection{Number Density}

Consider a general $2\to 2$ reaction with labels $1+2
\to 3+4$. In the MB approximation, from Eq.~\eqref{eq:collision}, the contribution of this reaction to the collision terms for the evolution of number densities is 
\beq
[{{C}}_n]^{12}_{34} & \equiv &
\Delta N\int\!d\Pi_1\!\int\!d\Pi_2
\;\mathcal{W}^{12}_{34}(s)\, 
f_1(T_1)f_2(T_2) \, ,
\label{eq:cnterm1}
\eeq
where $f_i(T_i) = (n_i/n_i^{eq})\exp(-E_i/T_i)$ and $\Delta N$ is the change in the relevant species number per reaction. This expression can be simplified by generalizing the approach of Refs.~\cite{Gondolo:1990dk,Edsjo:1997bg} to allow for each species to have an independent temperature $T_i$ and assuming massless particles. To do so, define new variables as
\beq
e_\pm = \frac{E_1}{T_1} \pm \frac{E_2}{T_2} \ ,
\qquad
\hat{s} = s/\left(T_1T_2\right) = \frac{1}{2}(e_+^2-e_-^2)(1-\cos\theta) \ ,
\eeq
where $\theta$ is the angle between $\vec{p}_1$ and $\vec{p}_2$. These vary over the ranges
\beq
\hat{s} \geq 0 \ ,
\qquad
e_+ \geq \sqrt{\hat{s}} \ ,
\qquad
|e_-| \leq \sqrt{e_+^2-\hat{s}} \ .
\eeq
Changing variables in Eq.~\eqref{eq:cnterm1} yields
\beq
[{{C}}_n]^{12}_{34} & = &
\Delta N\!\lrf{n_1n_2}{n_1^{eq}n_2^{eq}}
\frac{(T_1T_2)^2}{8(2\pi)^4}
\int\!d\hat{s}\!\int\!de_+\!\int\!de_-
\,\mathcal{W}^{12}_{34}(T_1T_2\hat{s})\;e^{-e_+}
\label{eq:cnterm2}\\
&=& \Delta N\!\lrf{n_1n_2}{n_1^{eq}n_2^{eq}}
\frac{(T_1T_2)^2}{8(2\pi)^4}
\int\!d\hat{s}\,\mathcal{W}^{12}_{34}(T_1T_2\hat{s})
\!\int\!de_+\;2\sqrt{e_+^2-\hat{s}}\,e^{-e_+}
\nnmb\\
&=& \Delta N\lrf{n_1n_2}{n_1^{eq}n_2^{eq}}\frac{\sqrt{T_1T_2}}{4(2\pi)^4}\int\!ds\,\mathcal{W}^{12}_{34}(s)\,\sqrt{s}\,K_1(\sqrt{s/T_1T_2}) \ ,
\nnmb
\eeq
where $K_1(z)$ is a modified Bessel function and in the last line we have used~\cite{NIST:DLMF}
\beq
(-\partial_z)^n[z^{-\nu}K_\nu(z)] = \frac{\mathbf{\Gamma}(1/2)}{2^\nu\mathbf{\Gamma}(\nu+1/2)}
\int_{1}^{\infty}\!dt\,t^n\,(t^2-1)^{\nu-1/2}\,e^{-zt} \ .
\label{eq:intbess}
\eeq
Note that the full collision term for the evolution of a given number density will receive contributions of this form from all reactions that change the number of the species, including the reverse reaction $3+4\to 1+2$.

The expression of Eq.~\eqref{eq:cnterm2} can be evaluated analytically when the scattering kernel $\mathcal{W}(s)$
is a polynomial in $s$ using the identity~\cite{NIST:DLMF}
\beq
\int_0^{\infty}\!dt\,t^{\mu-1}K_{\nu}(t) = 2^{\mu-2}\,
\mathbf{\Gamma}(\mu/2-\nu/2)\,\mathbf{\Gamma}(\mu/2+\nu/2) \ ,
\qquad
|\mathrm{Re}(\nu)|<\mathrm{Re}(\mu) \ .
\label{eq:besselmagic}
\eeq
As an example, applying this relation to the monomial form $\mathcal{W}^{12}_{34}(s) \to A_ks^k$, we obtain
\beq
[{{C}}_n]^{12}_{34} \ \to \ 
\lrf{n_1n_2}{n_1^{eq}n_2^{eq}}\frac{k!\,(k+1)!\,2^{2k}}{(2\pi)^4}\,
\,A_k(T_1T_2)^{k+2}
\eeq
This result coincides with the calculation of Ref.~\cite{Unwin:2014poa}. Let us also point out that the larger the power of $k$, the better the MB approximation used here is expected to be.

\subsection{Energy Density}

For energy transfer between sectors, we are primarily interested in reactions of the form $1+2\to 3+4$ where both initial states are in the visible sector at temperature $T$ and both final states are in a dark sector with an independent temperature $T_x$.
This reaction transfers energy from the visible sector to the dark sector with rate
\beq
[{{C}}_E]^{12}_{34} &\equiv&
\int\!d\Pi_1\!\int\!d\Pi_2
\;\mathcal{W}^{12}_{34}(s)\,(E_1+E_2)\,
f_1(T)f_2(T)
\label{eq:ceterm}\\
&=& \lrf{n_1n_2}{n_1^{eq}n_2^{eq}}\frac{T}{4(2\pi)^4}\int_0^\infty\!ds\,\mathcal{W}^{12}_{34}(s)\,{s}\,K_2(\sqrt{s}/T) \ .
\nnmb
\eeq
As before, we have generalized the analysis of Refs.~\cite{Gondolo:1990dk,Edsjo:1997bg} to obtain the second line. The corresponding expression for the reverse reaction would involve $T_x$ rather than $T$.

For the specific case of a monomial scattering kernel, $\mathcal{W}^{12}_{34} \to A_ks^k$, evaluating Eq.~\eqref{eq:ceterm} using the relation of Eq.~\eqref{eq:besselmagic} yields
\beq
[{{C}}_E]^{12}_{34} \ \to 
\lrf{n_1n_2}{n_1^{eq}n_2^{eq}}\frac{k!\,(k+2)!\,2^{2k+1}}{(2\pi)^4}\,
\,A_k\,T^{2k+5} \ .
\eeq

\section{\label{appx:Mcalc}Asymmetry Matrix Element Calculations}

In this appendix we elaborate on the calculation of asymmetric matrix elements and scattering kernels. As discussed in Sec.~\ref{subsec:overview_asymm}, these require both real and imaginary phases and arise at leading non-trivial order through the interference between tree-level and one-loop diagrams. In our scenario, the weak phases come from the couplings $y_{ia}$ and $\lambda_a$ while the strong phases arise from intermediate particles going on shell in loops. 

The diagrammatic topologies contributing to asymmetric matrix elements at leading order are shown in Fig.~\ref{fig:feynman_asym}.  Strong phases arising from loop diagrams can be obtained with unitarity cuts~\cite{Cutkosky:1960sp}. After cutting, each one-loop diagram can be associated with a pair of tree-level diagrams and their related weak phases. With this in mind, it is convenient to define the following products of couplings:
\beq
A_{s,ij} \ = \ A_{u,ij} &=& \sum_a\frac{y_{ia}M_ay_{ja}}{M_a^2}
\ \sim \ (\ell_iH\to\ell_j^\dagger H^\dagger)_{s,u} \ ,
\\
B_s \ = \ B_u &=& \sum_a\frac{\lambda_{a}M_a\lambda_{a}}{M_a^2}
\ \sim \ (\chi\phi\to\chi^\dagger\phi^\dagger)_{s,u} \ ,
\nnmb\\
C_i &=& \sum_a\frac{y_{ia}M_a\lambda_{a}}{M_a^2}
\ \sim \ (\ell_iH\to\chi^\dagger\phi^\dagger)_s\ ,
\nnmb\\
D_i &=& \sum_a \frac{y_{ia}\lambda_a^*}{M_a^2}
\ \sim \ (\ell_iH\to\chi\phi)_s \ ,
\nnmb
\eeq
where, on the right-hand side in each line, we indicate in which matrix element calculation the product appears.
The coefficients corresponding to the CP conjugates of each tree-level diagram is the same. For CP- or T- conjugate diagrams, the related coefficient should be complex conjugated. 

For any given process, only a subset of one-loop diagrams contribute to asymmetries. Depending on which particles run in the loop and where mass insertions occur, the product of complex couplings $\Im(c_0^*c_1)$ vanishes for many combinations. In terms of how asymmetries arise through loop topologies, there are three main classes: i) $\ell_i H \to \ell_j^\dagger H^\dagger$ and $\chi\phi\to\chi^\dagger\phi^\dagger$; ii) $\ell_iH \to \chi^\dagger\phi^\dagger$; iii) $\ell_i H\to \chi\phi$. We treat each of these in turn.

For $\ell_i H \to \ell_j^\dagger H^\dagger$ and $\chi\phi\to\chi^\dagger\phi^\dagger$ there are $s$- and $u$-channel diagrams at tree-level. One-loop asymmetries are generated by propagator loops on $s$-channel $N_a$ containing $\chi\phi$ for $\ell H\to\ell^\dagger H^\dagger$ and $\ell H$ for $\chi\phi \to \chi^\dagger\phi^\dagger$. There are contributions from Majorana mass insertions on either side of the loop. At leading order in an expansion in powers of Mandelstam variables, we find
\beq
\abs*{\Delta\mathcal{M}^{\ell_iH}_{\ell_j^\dagger H^\dagger}}^2 = -\frac{st}{4\pi}\,
\Im\!\left[\left(2A_{s,ij}^{*}+A_{u,ij}^{*}\right)\!\left(C_iD_j+D_iC_j\right)\right]
\eeq
and
\beq
\abs*{\Delta\mathcal{M}^{\chi\phi}_{\chi^\dagger\phi^\dagger}}^2 = -\frac{st}{4\pi}\,
\mathrm{Im}\!\left[\left(B_s^{*}+B_u^{*}\right)\sum_k\left(C_kD_k^*+D_k^*C_k\right)
\right] \ . 
\label{eq:masym1}
\eeq
The relative factors of two in some of terms correspond to differences in SU(2)$_L$ contractions.

The second class of asymmetries corresponds to the reaction $\ell_i H\to \chi^\dagger \phi^\dagger$ and its conjugates. There is now only a single $s$-channel contribution at tree-level. The loops relevant to asymmetries are vertex loops on either side of the diagram, and propagator loops on an $s$-channel $N_a$ leg containing $\ell_kH$ and a Majorana mass insertion on the $\ell_i H$ side or with $\chi\phi$ and a mass insertion on the $\chi\phi$ side. We find the asymmetric matrix element
\beq
\abs*{\Delta\mathcal{M}^{\ell_i H}_{\chi^\dagger\phi^\dagger}}^2 = -\frac{st}{4\pi}\,
\Im\!\left[C_i^*\!\left(\sum_kA_{u,ik}D_k^* + D_iB_u + 2\sum_kA_{s,ik}D_k^* + D_iB_{s}\right)\right] \ .
\label{eq:masym2}
\eeq
In the last bracket above, the first two terms correspond to the vertex loops and the second two to the propagator loops.

The last class of processes supporting an asymmetry is $\ell_i H\to \chi \phi$. This reaction preserves the combined charge $L+L_x$ and requires an even number of Majorana neutrino mass insertions. To generate asymmetries, the single $s$-channel contribution at tree level interferes with vertex loops at both sides of the diagram as well as propagator loops containing $\chi\phi$ and $\ell_kH$ on an $s$-channel Majorana neutrino line. The asymmetry is
\beq
\abs*{\Delta\mathcal{M}^{\ell_i H}_{\chi\phi}}^2 = -\frac{su}{4\pi}\,
\Im\!\left[D_i^*\!\left(\sum_kA_{u,ik}C_k^* + C_iB_{u}^*
+2\sum_k{A_{s,ik}}C_k^* + C_iB_s^{*}
\right)\right] \ .
\label{eq:masym3}
\eeq
As above, in the last bracket the first two terms correspond to vertex loops and the last two terms are propagator loops.

Integrating these expressions over the final state phase spaces yields the related asymmetric scattering kernels.  This produces $-st\to s^2/16\pi$ and $-su\to s^2/16\pi$, and thus the leading terms in all the scattering kernels are proportional to $s^2$. These can be shown to satisfy the CPT/unitarity constraint of Eq.~\eqref{eq:cptu}. Specializing to $n_g=3$ lepton generations with $y_{ia}=y_a$, we obtain the asymmetry coefficients $\nlodelta$ of Eqs.~\eqref{eq:d1a}--\eqref{eq:d1d}.


\section{Source and Wash-out Collision Term Calculations}
\label{appx:fullBE}

We turn now to elaborating on the calculation of the collision terms $C_Q$ for the charges ${Q=B-L,\,L_x}$ defined in Eq.~\eqref{eq:boltzq}. In our scenario, these are dominated by $2\to 2$ reactions of the form $1+2\to 3+4$. For the discussion here, it is convenient to classify such reactions as self-conjugate, satisfying $(34)=(12)^\dagger$, and non-self-conjugate otherwise. The total contribution to $C_Q$ from a self-conjugate reaction involves only $1+2\to {1}^\dagger+{2}^\dagger$ and ${1}^\dagger+{2}^\dagger\to 1+2$. For a non-self-conjugate reaction, there are also independent reverse reactions giving four contributions: $1+2\to 3+4$, ${1}^\dagger+{2}^\dagger\to {3}^\dagger+{4}^\dagger$, $3+4\to 1+2$, and ${3}^\dagger+{4}^\dagger\to {1}^\dagger+{2}^\dagger$.

Consider first a self-conjugate reaction $1+2\to 3+4$, with $3=1^\dagger$ and $4=2^\dagger$, that changes the charge $Q$ by the amount $\Delta Q$. The total contribution to $C_Q$ from Eqs.~(\ref{eq:collision},\ref{eq:cnterm1}) is
\beq
\Delta C_Q\big/\Delta Q &=& \int\!d\Pi_1\!\int\!d\Pi_2\left(\mathcal{W}^{12}_{34}\,f_1f_2 -\mathcal{W}^{{1^\dagger2^\dagger}}_{{3^\dagger4^\dagger}}\,f_{1^\dagger}f_{{2}^\dagger}\right)
\ ,
\eeq
where 
\beq
f_{i,{i}^\dagger} = e^{\pm\mu_i/T_i}e^{-E_i/T_i} \ \equiv \ e^{\pm\mu_i/T_i}f_{i}^0
\ ,
\eeq
with $T_i$ being the relevant temperature of the $i$-th species and $\mu_i$ its chemical potential. Since we expect small asymmetries, it is a good approximation to expand to linear order in the chemical potentials. We find
\beq
\Delta C_Q\big/\Delta Q &=& \int\!d\Pi_1\!\int\!d\Pi_2
\left[
\Delta\mathcal{W}^{12}_{34}+\Sigma\mathcal{W}^{12}_{34}\left(\frac{\mu_1}{T_1}+\frac{\mu_2}{T_2}\right)
\right]f_1^0f_2^0
\ .
\label{eq:delcqself}
\eeq
Referring back to Eq.~\eqref{eq:cqsplit}, we identify the first term above with part of the source term $S_Q$ and the second term with a portion of the wash-out term $W_Q$.

The contribution to the charge collision term $C_Q$ from a non-self-conjugate reaction $1+2\to 3+4$ follows similarly, but now with additional pieces coming from the independent reverse reactions. These can be simplified using the effective masslessness of all initial and final states, CPT, and our definitions of symmetric and asymmetric matrix elements. Together, these imply $\Sigma\mathcal{W}^{34}_{12}(s) = \Sigma\mathcal{W}^{12}_{34}(s)$ and $\Delta\mathcal{W}^{34}_{12}(s) = -\Delta\mathcal{W}^{12}_{34}(s)$. Assembling all the terms gives the final result
\beq
\Delta C_Q\big/\Delta Q &=& \int\!d\Pi_1\!\int\!d\Pi_2
\Bigg(
\Delta\mathcal{W}^{12}_{34}\left(f_1^0f_2^0+f_3^0f_4^0\right)
\label{eq:delcqnonself}
\\
&&
\hspace{1cm}
\ + \ \Sigma\mathcal{W}^{12}_{34}\bigg[
\left(\frac{\mu_1}{T_1}+\frac{\mu_2}{T_2}\right)f_1^0f_2^0
-\left(\frac{\mu_3}{T_3}+\frac{\mu_4}{T_4}\right)f_3^0f_4^0
\bigg]
\Bigg)
\ ,
\nnmb
\eeq
where here $f_3^0=\exp(-E_1/T_3)$ and $f_4^0 = \exp(-E_2/T_4)$ are functions of the momenta being integrated over and the scattering kernels are functions of $s=(p_1+p_2)^2$. As for the self-conjugate case, the first term depending on $\Delta\mathcal{W}^{12}_{34}$ contributes to the source term for $Q$ while the second that depends on $\Sigma\mathcal{W}^{12}_{34}$ contributes to the wash-out term.

The full charge collision term $C_Q$ is the sum of contributions from all relevant reactions that change charge $Q$. For our scenario, these are collected in Tab.~\ref{tab:transfer}. Collecting them all and relating the species chemical potentials to the charge chemical potentials using Eqs.~(\ref{eq:musm1},\ref{eq:muhid1}), we obtain the source and wash-out terms collected in Sec.~\ref{subsec:BE_TRH}. 

To give more insight into how the general results here translate into the specific expressions in Sec.~\ref{subsec:BE_TRH}, let us now present two examples. First, we derive the source term for $B-L$. 
This example also demonstrates how CPT and unitarity resolve the seemingly curious feature of Eqs.~(\ref{eq:delcqself},\ref{eq:delcqnonself}) that the source contributions from the individual contributions do not vanish in equilibrium, corresponding to $T_1=T_2=T_3=T_4$. The reactions contributing to the $B-L$ source are $\ell_iH\to \ell^{\dagger}H^{\dagger}$~($\Delta Q = 2$), $\ell_iH\to \chi\phi$~($\Delta Q = 1$), and $\ell_i H\to\chi^\dagger\phi^\dagger$~($\Delta Q = 1$). Summing over them, we obtain
\beq
S_Q &=& \int\!d\Pi_1\!\int\!d\Pi_2\;\sum_i\!\left[
\sum_{j}\Delta\mathcal{W}^{\ell_iH}_{\ell_j^\dagger H^\dagger}\,2\,f_\ell^0 f_H^0
+\Big(\Delta\mathcal{W}^{\ell_iH}_{\chi\phi} +\Delta\mathcal{W}^{\ell_iH}_{\chi^\dagger\phi^\dagger}\Big)
\Big(f_\ell^0 f_H^0+f_\chi^0f_\phi^0\Big)
\right]~~~~\\
&=&
\sum_{i,j}\int\!d\Pi_1\!\int\!d\Pi_2\;
\Delta\mathcal{W}^{\ell_iH}_{\ell_j^\dagger H^\dagger}
\Big(f_\ell^0 f_H^0-f_\chi^0f_\phi^0\Big)
\nnmb
\eeq
In going from the first to the second line, we have applied the CPT and unitarity relation of Eq.~\eqref{eq:cptu} (as realized explicitly in Eqs.~(\ref{eq:cptu1},\ref{eq:cptu2})). Upon integrating over the initial state phase space, we reproduce Eq.~\eqref{eq:sourcebl}. We also see that, in contrast to the individual contributions to the source, the full source term vanishes in equilibrium when $T_x=T$.

As a second example, let us compute the contributions to the $B-L$ and $L_x$ wash-out terms from the reaction $\ell_iH\to \chi^\dagger\phi^\dagger$, which has $\Delta(B-L)=+1$ and $\Delta L_x=-1$. Applying our general result, we get
\beq
\Delta W_{Q}/\Delta Q &=& \int\!d\Pi_1\!\int\!d\Pi_2\;\Sigma\mathcal{W}^{\ell_i H}_{\chi^\dagger\phi^\dagger}\left[\bigg(\frac{\mu_\ell+\mu_H}{T}\bigg){f_\ell^0}f_H^0 - \bigg(\frac{-\mu_\chi-\mu_\phi}{T_x}\bigg)f_\chi^0f_\phi^0
\right] \\
&=&
-\frac{66}{79}\frac{n_{B-L}}{T^3}\int\!d\Pi_1\!\int\!d\Pi_2\;\Sigma\mathcal{W}^{\ell_i H}_{\chi^\dagger\phi^\dagger}\,{f_\ell^0}f_H^0
\nnmb\\
&&
~~~~~+ 6\frac{n_{L_x}}{T_x^3}\int\!d\Pi_1\!\int\!d\Pi_2\;\Sigma\mathcal{W}^{\ell_i H}_{\chi^\dagger\phi^\dagger}\,{f_\chi^0}f_\phi^0
\nnmb
\eeq
Here, we have used the relations between chemical potentials and charge densities obtained in Sec.~\ref{subsec:chemicals}. Integrating over these expressions over phase space reproduces the corresponding parts of Eq.~\eqref{eq:wash-out}.

\section{Transfer with Non-Instantaneous Reheating\label{appx:reheat}}

{

In our primary analysis above we assumed instantaneous reheating after inflation to a temperature $\Trh < M_1$. Here we consider the impact of non-instantaneous reheating on our scenario. Based on our estimates, we argue that most of the results obtained with instantaneous reheating also apply to motivated non-instantaneous cases as long as the ratio $M_1/\Trh$ is large enough.

As a simple model for the reheating process, we consider a reheating period between the inflation and radiation eras during which the total energy density of the universe is dominated by a scalar field (such as the inflaton) oscillating in a quadratic potential as it decays exclusively to light SM particles. The averaged energy density of these oscillations redshifts like matter, ${\rho_o\propto a^{-3}}$. After a brief initial transient, the subdominant visible radiation component reaches a maximum and then redshifts as $\rho_r\propto T^4\propto a^{-3/2}$. Reheating is completed when the radiation density becomes dominant at temperature $\Trh$. For the purposes of our estimates, a sufficient parametrization of this behavior is
\beq
\Hub \simeq \left\{
\begin{array}{lcc}
\Hrh\,a^{-3/2}&;&T>\Trh\\
\Hrh\,a^{-2}&;&T<\Trh
\end{array}
\right. \ ,
\qquad
T \simeq \left\{
\begin{array}{lcc}
\Trh\,a^{-3/8}&;&T>\Trh\\
\Trh\,a^{-1}&;&T<\Trh
\end{array}\right. \ ,
\eeq
where we have normalized $a=1$ at reheating and $\Hrh \sim \Trh^2/\mpl$. We also define ${\xrh = M_1/\Trh}$ and assume that it is larger than unity.

Consider now the transfer of energy density from the visible sector to the dark sector during and after reheating. The dominant reactions contributing to the transfer are $2\to2$ processes such as $\ell H\to \chi^\dagger\phi^\dagger$ mediated by an off-shell $N_a$, together with $2\to 1$ production $\ell H\to N_a$ followed by the $1\to 2$ decay $N_a\to \chi\phi$. To compute the total amount of energy transferred, it is helpful to consider the contributions from three disjoint regions in the evolution: i) $a\in [a_i,a_M]$, ii) $a\in [a_M,1]$, iii) $a\in [1,a_f]$, where $a_i$ is some initial scale factor after inflation, $a_M$ is the scale factor at which $T = M_1$, and $a_f$ is some final scale factor after the end of reheating. Regions i) and ii) occur during the reheating process, while region iii) is the evolution after reheating has completed. Note that we implicitly allow the maximum temperature achieved during reheating to exceed $M_1$. For simplicity, we also consider only a single flavor of massive neutrino $N_1$.

Let us estimate first the transfer of energy to the dark sector energy from $2\to 2$ reactions. The contribution from region i) with $a < a_M$ can be shown to be subleading, so we discuss only the portions from regions ii) and iii) here. When $a > a_M$, the leading part of the energy transfer collision term is
\beq
C_E \ \simeq \ \frac{A}{M_1^{2k}}\left(T^{2k+5}-T_x^{2k+5}\right) \ ,
\eeq
with $k=1$ and $A \simeq 3M_1^2[\mathcal{S}_1]^{\ell  H}_{\chi^\dagger\phi^{\dagger}}\big/\pi^4 $. Applying the form to region ii), neglecting reverse reactions we find the upper bound 
\beq
\Delta(a^4\rho_x) &\lesssim&
\int_{a_M}^1\!da\;\frac{a^4}{a \Hub}\,A\,\frac{T^{2k+5}}{M_1^{2k}}\\
&\simeq& \Trh^4\lrf{AM_1}{\Hub_{\rm RH}}\xrh^{-(2k+1)}\Big(1 - a_M^{(29-6k)/8}\Big)\frac{1}{29-6k}  \ .
\nnmb
\eeq
This contribution is dominated by $T\sim \Trh$ for $k < 29/6$, as we have here. Similarly, energy transfer from the visible sector to the dark sector in region iii) after reheating is
\beq
\Delta(a^4\rho_x) &\lesssim&
\int_1^{a_f}\!da\;\frac{{a}^4}{a\Hub}\,A\,\frac{T^{2k+5}}{M_1^{2k}}\\
&\simeq& \Trh^4\lrf{AM_1}{\Hrh}\xrh^{-(2k+1)}\Big(1 - a_f^{-(2k-1)}\Big)\frac{1}{2k-1} \ .
\nnmb
\eeq
This result matches our previous calculation in Sec~\ref{subsec:heat_transfer},
and is dominated by reactions near $T\sim \Trh$. Comparing the contributions from regions ii) and iii), we see that they are of the same order. A similar conclusion holds for asymmetry generation and washout, which is dominated in our scenario by operators with $k=1$~(washout) and $k=2$~(asymmetry sources).

We turn next to the transfer of energy to the dark sector from the direct production and decay of the massive neutrinos. The relevant energy transfer collision term for total decay width $\Gamma \gg \Hub$ of the massive neutrino $N_1$ is
\beq
C_E \ \simeq \ M_1\frac{\Gamma_{N_1\to\ell H}\Gamma_{N_1\to\chi\phi}}{\Gamma}\Big[\bar{n}_1(T)-\bar{n}_1(T_x)\Big] \ ,
\eeq
where the partial widths are listed in Eq.~\eqref{eq:gamman}.
Starting with region i), we have 
\beq
\Delta(a^4\rho_x) &\lesssim& M_1\frac{\Gamma_{N_1\to\ell H}\Gamma_{N_1\to\chi\phi}}{\Gamma}\,\int_{a_i}^{a_M}\!da\;\frac{a^4}{a\Hub}\frac{2(3/4)\zeta(3)}{\pi^2}T^3\\
&\simeq& \Trh^4\,\frac{12\zeta(3)}{35\pi^2}\,\frac{\Gamma_{N_1\to\ell H}\Gamma_{N_1\to\chi\phi}}{\Gamma\,\Hrh}\,\xrh^{-32/3} \ .
\nnmb
\eeq
From region ii) we find
\beq
\Delta(a^4\rho_x) &\lesssim& M_1\frac{\Gamma_{N_1\to\ell H}\Gamma_{N_1\to\chi\phi}}{\Gamma}\,\int_{a_M}^{a_1}\!da\;\frac{a^4}{a\Hub}2\lrf{M_1T}{2\pi}^{3/2}e^{-M_1/T}\\
&\simeq& \Trh^4\,\frac{16}{3(2\pi)^{3/2}}\,\frac{\Gamma_{N_1\to\ell H}\Gamma_{N_1\to\chi\phi}}{\Gamma\,\Hrh}\,\xrh^{-32/3}\,\mathbf{\Gamma}(73/6+1) \ .
\nnmb
\eeq
Note that in the second line above we have used
\beq
\int_1^{\xrh}\!du\;u^{73/6}e^{-u} \ \lesssim \ \mathbf{\Gamma}(73/6+1) \ ,
\eeq
where $\mathbf{\Gamma}(z)$ is the gamma function
and this expression is an approximate equality for $\xrh \gg 73/6$. Finally, in region iii) we obtain
\beq
\Delta(a^4\rho_x) &\lesssim& M_1\,\frac{\Gamma_{N_1\to\ell H}\Gamma_{N_1\to\chi\phi}}{\Gamma}\,\int_{1}^{a_f}\!da\;\frac{a^4}{a\Hub}\,2\!\lrf{M_1T}{2\pi}^{\!3/2}\!e^{-M_1/T}\\
&\simeq& \Trh^4\,\frac{2}{(2\pi)^{3/2}}\,\frac{\Gamma_{N_1\to\ell H}\Gamma_{N_1\to\chi\phi}}{\Gamma\,\Hrh}\,\xrh^{3/2}\,e^{-\xrh} 
\ .\nnmb
\eeq
Examining the contributions from the three regions, we find that region ii) is dominant.

We are now equipped to compare the relative contributions from $2\to 2$ and $2\to 1$ processes to the dark sector energy density. 
Examining the scalings of the various contributions with $\xrh$, the dominant source of dark sector energy will come from $2\to 2$ reactions near reheating if $\xrh$ is big enough. Evaluating numerically, and noting that $AM_1 \sim |\lambda_1|^2\Gamma_{N_1\to\ell H}\Gamma_{N_1\to\chi\phi}/\Gamma$ for $|\lambda|^2\gg |y|^2$, we find that $2\to 2$ reactions near $\Trh$ dominate the energy transfer to the  dark sector when $\xrh \gtrsim 20$ for $|\lambda_1|= 1.0$ and when $\xrh \gtrsim 35$ for $|\lambda_1|=0.1$. Also, fixing $\xrh = 30$ we find that $2\to 2$ reactions near $\Trh$ are the leading source of energy transfer to the dark sector throughout the parameter region where the mechanism can generate the observed baryon abundance.
Based on these calculations, we conclude that replacing our previous assumption of instantaneous reheating with a more gradual reheating process as modeled above our results for energy transfer to the dark sector would not change in a significantly qualitative way for most of the parameter ranges we consider. Nearly identical arguments can be applied to asymmetry generation through the decays of massive neutrinos created prior to reheating; their contribution is subleading to that from $2\to 2$ processes after reheating provided $\xrh$ is large enough.
}


\end{spacing}

\newpage
\bibliography{freezein_BG}
\bibliographystyle{utphys}

\end{document}